\documentclass[sn-basic]{sn-jnl}

\usepackage{graphicx}%
\usepackage{multirow}%
\usepackage{amsmath,amssymb,amsfonts}%
\usepackage{amsthm}%
\usepackage{mathrsfs}%
\usepackage[title]{appendix}%
\usepackage{xcolor}%
\usepackage{textcomp}%
\usepackage{manyfoot}%
\usepackage{booktabs}%
\usepackage{algorithm}%
\usepackage{algorithmicx}%
\usepackage{algpseudocode}%
\usepackage{listings}%
\DeclareMathOperator*{\argmax}{argmax}   
\DeclareMathOperator*{\argmin}{argmin}   
\usepackage{titlesec}

\theoremstyle{thmstyleone}%
\theoremstyle{thmstyletwo}%

\theoremstyle{thmstylethree}%
\renewcommand{\appendixname}{Supplementary Material}

\begin{document}

\title[]{Lower-dimensional posterior density and cluster summaries for overparameterized Bayesian models}


\author*[1]{\fnm{Henrique} \sur{Bolfarine}}\email{henrique.bolfarine@austin.utexas.edu}

\author[2]{\fnm{Hedibert} \sur{F. Lopes}}

\author[1]{\fnm{Carlos} \sur{M. Carvalho}}

\affil*[1]{\orgdiv{McCombs School of Business}, \orgname{The University of Texas at Austin}, \orgaddress{ \city{Austin}, \state{Texas}, \country{USA}}}

\affil[2]{\orgname{Insper Institute of Education and Research}, \orgaddress{\state{S\~ao Paulo}, \country{Brazil}}}




\abstract{The usefulness of Bayesian models for density and cluster estimation is well established across multiple literatures. However, there is still a known tension between the use of simpler, more interpretable models and more flexible, complex ones. In this paper, we propose a novel method that integrates these two approaches by projecting the fit of a flexible, overparameterized model onto a lower-dimensional parametric surrogate, which serves as a summary. This process increases interpretability while preserving most of the fit of the original model. Our approach involves three main steps. First, we fit the data using nonparametric or overparameterized models. Second, we project the posterior predictive distribution of the original model onto a sequence of parametric summary point estimates with varying dimensions using a decision-theoretic approach. Finally, given the parametric summary estimate, obtained in the second step, that best approximates the original model, we construct uncertainty quantification for this summary by projecting the original posterior distribution. We demonstrate the effectiveness of our method for generating summaries for both nonparametric and overparameterized models, delivering both point estimates and uncertainty quantification for density and cluster summaries across synthetic and real datasets.}

\keywords{Model summarization, Decision theory, Density estimation, Clustering, Finite mixture models.}



\maketitle

\section{Introduction}\label{section1}

Within the context of Bayesian density estimation and clustering, a number of well-established methods have been developed, with applications spanning areas such as medicine, biology, and engineering \citep[see][and references therein]{fruhwirth2019handbook, mclachlan2019finite, wade2023bayesian}. Nonetheless, as is well known, there remains a tension between flexibility and interpretability in these models. For example, parametric approaches, such as finite mixture models \citep[][]{mclachlan2000finite, fruhwirth2006finite}, excel in interpretability, as they represent complex distributions as finite weighted sums of known distributions. More formally, a $d$-dimensional observation $\boldsymbol{y}_i \in \mathbb{R}^d$, originates from a finite mixture model with $K\in \mathbb{Z}_{+}$ components if it is drawn from a density of the form 
\begin{equation} \label{equation1}
	f(\boldsymbol{y}_i|\boldsymbol{\theta}) = \sum_{q = 1}^K \omega_q f(\boldsymbol{y}_i|\boldsymbol{\theta}_q),
\end{equation} 
where $\omega = (\omega_1,\dots,\omega_K)$ is a vector of positive weights such that $\sum_{q = 1}^K \omega_q = 1$, and $f(\cdot | \boldsymbol{\theta}_q)$ denotes a component specific kernel density, with parameters $\boldsymbol{\theta}_1, \dots,\boldsymbol{\theta}_K$, and $\boldsymbol{\theta} = (\omega_1,\dots,\omega_K;\boldsymbol{\theta}_1,\dots,\boldsymbol{\theta}_K)$.  Gaussian kernels are a common choice for $f(\cdot | \boldsymbol{\theta}_q)$, resulting in Gaussian mixture models (GMM).


Although finite mixture models are straightforward to interpret due to their modular nature, they may produce biased estimates as a result of restrictive modeling assumptions, thus failing to fully capture the complexity of the underlying distribution.

In contrast, nonparametric models, such as the Dirichlet process mixtures \citep[][]{ferguson1983bayesian, escobar1995bayesian} and random Bernstein polynomials \citep[BP;][]{petrone1999random, petrone1999bayesian}, allow for a more flexible and accurate representation of the data. In Dirichlet process mixture models, one of the most popular approaches, $\boldsymbol{y}_i$ is an observation from $f(\cdot|\boldsymbol{\theta}_i)$, where $\boldsymbol{\theta}_i$ is drawn from a discrete distribution $G$, itself defined by a Dirichlet process $DP(\alpha, G_0)$, with concentration parameter $\alpha > 0$ and base distribution $G_0$. In this setting, the distribution $G$ serves as the mixing measure, $G = \sum_{i \geq 1} \omega_i \delta_{\boldsymbol{\theta}_i}$, where $\boldsymbol{\theta}_1, \boldsymbol{\theta}_2, \dots$ are a sequence of random variables drawn from $G_0$, $\omega_1, \omega_2, \dots$ are random weights satisfying $\sum_{i \geq 1} \omega_i = 1$, and  $\delta_{\boldsymbol{\theta}_i}$ is the Dirac measure \citep{muller2015bayesian}. This formulation highlights the flexibility of Dirichlet process mixture models as there is no restriction on the number of parameters. 

However, Dirichlet process mixtures and similar approaches can generate models with an excessive number of components, leading to challenges in interpretability and biased estimates \citep{miller2013simple, miller2014inconsistency}.

In this paper, we propose a method that addresses the aforementioned issues by connecting simple, interpretable models to complex, nonparametric approaches. In a nutshell, we take advantage of the flexibility and predictive power of overparameterized and nonparametric models for density and cluster estimation and project the resulting fit onto a lower-dimensional parametric summary, serving as a surrogate. As a result, we obtain an interpretable, parsimonious summary that still preserves much of the fit and predictive accuracy of the original model. Furthermore, our approach allows us to measure the trade-off involved in performing such a projection.

The proposed method consists of three steps. In the first step, we estimate the model by adopting a flexible prior, which can be either nonparametric or overparameterized. In the second step, using a Bayesian decision-theoretic approach, \citep{berger2013statistical, hanh-carvalho-2015}, we project the posterior predictive distribution of the original model onto surrogates with different parametric dimensions. This process yields a sequence of summary point estimates of varying dimensions that capture different levels of complexity relative to the original model. To select the lower-dimensional summary estimate that best approximates the fit of the reference model, we propose a strategy that measures the divergence between the summary estimates and the original posterior predictive distribution. Finally, in the third step, given the summary identified in step two, we construct uncertainty quantification by projecting the full posterior distribution, generating a lower-dimensional posterior summary. This posterior summary is useful for inference and is well suited to addressing various questions about the data and the underlying distribution. Moreover, it remains valid in the sense that the data are utilized only once, conditional on the model's posterior distribution.

A key feature of our approach is that it imposes no restrictions on the models to which it can be applied. In this paper, we implement our method across different scenarios, including univariate and multivariate models, aiming to produce density and cluster summaries. Specifically, we consider nonparametric models such as random Bernstein polynomials and Dirichlet process mixtures, as well as methods that incorporate uncertainty in the number of mixture components, including mixtures of finite mixtures \citep[][]{richardson1997bayesian, miller2018mixture} and sparse overparameterized finite mixture models \citep[][]{malsiner2016model}.

\subsection{Motivating Example}

\begin{figure*}[t!]
	\centering
	\includegraphics[width=1\textwidth]{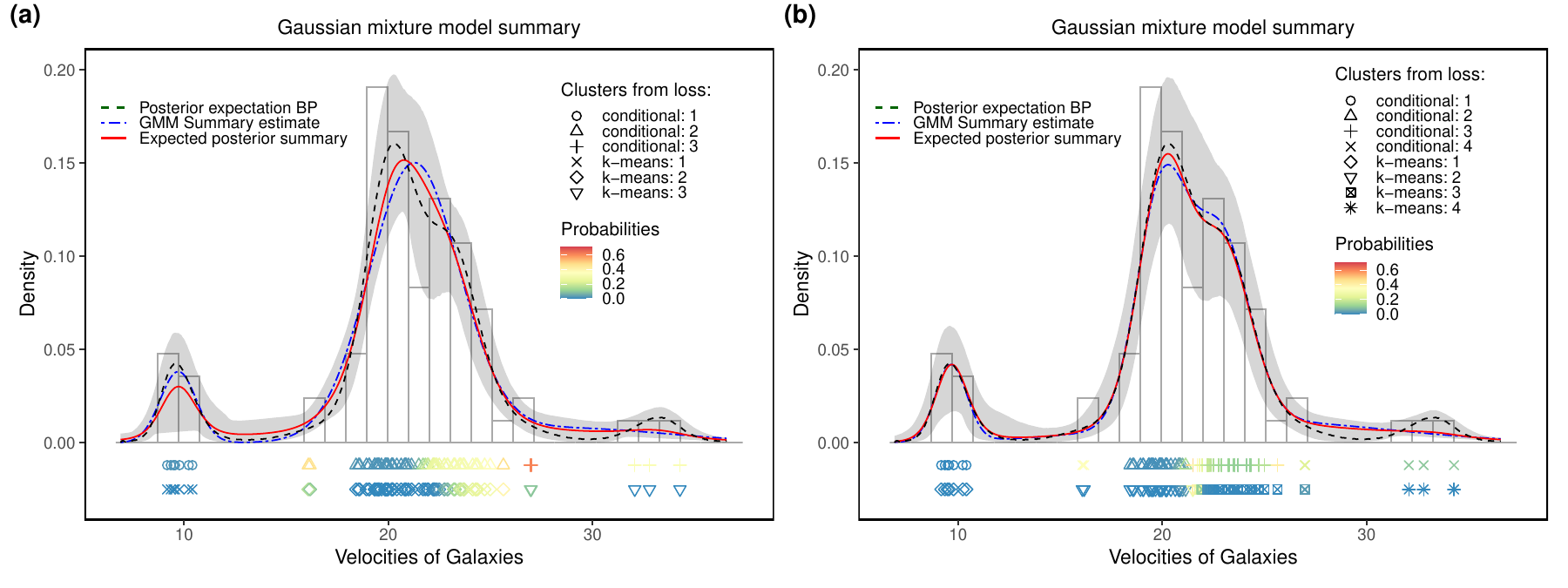}
	\caption{This figure displays the paper's main contributions. Figures (a) and (b) show histograms of data consisting of 82 galaxy velocities measured in \( 10^3 \text{ km/sec} \). A random Bernstein polynomial model was fitted to the data, where the black dashed line indicates the posterior expected density under this model. We applied the methodology presented in this paper to project the posterior distribution of the original model onto a surrogate. In this case, the summary model is a finite GMM with three components in (a) and four components in (b), shown as blue dot-dashed lines. From this projection, we additionally obtained a 95\% credible interval of the summary (gray ribbon), which encompasses the expected posterior distribution of the BP model in Figures (a) and (b). The red line represents the expected value of the projected posterior summary. Moreover, as shown in plots (a) and (b), our approach allows derivation of both optimal cluster allocations and posterior uncertainty quantification under conditional probabilities (row one) and k-means loss (row two).}\label{BP_galaxy_all}
\end{figure*}

In Figure \ref{BP_galaxy_all}, we can observe the paper's main contribution. It displays the well-known galaxy data consisting of 82 galaxy velocities, measured in $10^3$ kilometers per second, sampled from the Corona Borealis \citep{roeder1990density}. We fitted this data using a random Bernstein polynomial model \citep{petrone1999random, petrone1999bayesian}, implemented with the \texttt{DPpackage} \citep{JSSv040i05}. This model is known for producing accurate density estimates \citep{petrone2002consistency}; however, it is not suited for clustering as it tends to generate an excessive amount of groups. Details of the random Bernstein polynomial model and the configurations used to obtain this estimate are given in the Supplementary Material \ref{secA1}. The black dashed line in Figure \ref{BP_galaxy_all}, plots (a) and (b), is the expected posterior density generated under the BP model. Using a decision-theoretic approach, we projected the posterior distribution from the original model onto GMM summary point estimates, resulting in surrogate models with three and four components, which provided the best approximations. These summaries are shown in Figure \ref{BP_galaxy_all}, plots (a) and (b), as the blue dot-dashed line. Furthermore, we were able to project the full original posterior to the lower-dimensional GMM surrogate, resulting in 95\% credible intervals, seen in the gray ribbons, and the corresponding expected summary posterior density, shown in red, in both plots. Using the same posterior and different cluster allocation loss functions, we generated groups under the conditional probability allocation from the GMM summary and the k-means loss \citep{bishop2006pattern}, with uncertainty estimates for their allocation, shown in the first and second rows of Figure \ref{BP_galaxy_all}, plots (a) and (b), respectively. More details on this result are given in Section \ref{section6}.

\subsection{Related Work}

Our proposed method builds upon a rich literature using decision analysis for posterior summarization. \cite{maceachern2001decision} and \cite{gutierrez2005statistical} used a similar approach to summarize nonparametric regression models. \cite{hanh-carvalho-2015} used decision analysis to perform variable selection in regression models under sparse priors, extending the ideas from \cite{lindley1968choice}. Similar techniques were adopted for seemingly unrelated regressions \citep{puelz2017variable}, graphical models \citep{bashir2019post}, time-varying parameters \citep{huber2021inducing}, and factor analysis \citep{bolfarine2024decoupling}. \cite{woody2021model} proposed projecting the posterior distribution of Gaussian processes to a lower-dimensional posterior summary equipped with uncertainty quantification. This approach has since been successfully applied to create posterior summaries for causal estimates for growth mindset intervention, as seen in \cite{yeager2019national}. \cite{fisher2024bayesian} applied a similar procedure to estimate heterogeneous treatment effects in randomized trials with noncompliance. Recently, \cite{buch2024bayesian} introduced a method for estimating level-set clustering using nonparametric surrogates. Compared to this approach, our method provides a more flexible and modular alternative with the possibility of using different loss functions and summary classes for both density and cluster estimation.

This paper is structured as follows. In Section \ref{section2}, we provide an overview of the decision-theoretic approach, in which we discuss the choice of the loss function and the class of surrogates to generate the sequence of summary point estimates. In Section \ref{section3}, given the dimension of the summary with the best approximation identified in step two, we project the posterior distribution of the original model to generate a posterior summary, which is equipped with uncertainty quantification. In Section \ref{section5}, we extend the procedure to provide both optimal estimates and uncertainty quantification for cluster summaries. In Section \ref{section6}, we apply our method to synthetic and real data under different models and end with a discussion in Section \ref{section7}.

\section{Lower-Dimensional Summary Estimates for Densities}\label{section2}

\subsection{Decision-Theoretic Framework}\label{section21}

Consider $N$ independent observations $\boldsymbol{y} = (\boldsymbol{y}_1, \dots, \boldsymbol{y}_N)^{T}$, where $\boldsymbol{y}_i \in \mathbb{R}^d$. We assume that each $\boldsymbol{y}_i$ is generated from an unknown probability distribution $\mathbb{P}_{\boldsymbol{\theta}}$, indexed by a parameter $\boldsymbol{\theta} \in \boldsymbol{\Theta}$. Here, the goal is to learn $f_{\boldsymbol{\theta}}(\cdot):= f(\cdot|\boldsymbol{\theta})$, the density associated with $\mathbb{P}_{\boldsymbol{\theta}}$, based on the observed data. 

In the first step of our method, we apply standard Bayesian procedures for inference given a prior distribution $p(\boldsymbol{\theta})$ and a model for $f_{\boldsymbol{\theta}}$. As a result, we obtain the posterior distribution $p(\boldsymbol{\theta} | \boldsymbol{y})\propto f(\boldsymbol{y}|\boldsymbol{\theta})p(\boldsymbol{\theta})$ which serves as the basis for all inferences about $f_{\boldsymbol{\theta}}$. We refer to $p(\boldsymbol{\theta}|\boldsymbol{y})$, as the original posterior distribution. A natural choice for the prior $p(\boldsymbol{\theta})$ is a nonparametric or overparameterized family of distributions which better captures the complexity of the data. However, as discussed in Section \ref{section1}, the posterior distribution under such priors can lead to overfitting and misleading results.

To address this challenge, in the second step, we introduce a decision-theoretic approach \citep{berger2013statistical, hanh-carvalho-2015}, with the goal of projecting the posterior predictive distribution yielded by the original model, given by $\tilde{f}(\boldsymbol{\tilde{y}}) := f(\boldsymbol{y}_{N+1} | \boldsymbol{y})$, where  $f(\boldsymbol{y}_{N+1} | \boldsymbol{y})  = \int f(\boldsymbol{y}_{N+1} | \boldsymbol{\theta}) p(\boldsymbol{\theta} | \boldsymbol{y}) d\boldsymbol{\theta}$, onto a lower-dimensional surrogate, which serves as a summary. We refer to $\tilde{f}$ as the reference model. The surrogate density is defined as $g_{\boldsymbol{\gamma}}\in {\cal G}_{\boldsymbol{\gamma}}$, with  $g_{\boldsymbol{\gamma}}(\cdot) := g ( \cdot |\boldsymbol{\gamma})$, and where  ${\cal G}_{\boldsymbol{\gamma}} := \{ g ( \cdot |\boldsymbol{\gamma}); \boldsymbol{\gamma} \in \boldsymbol{\Gamma} \}$ is a class of density functions that share the same support as $\boldsymbol{y}_i$. In this case, the parameter of the surrogate, $\boldsymbol{\gamma}$, serves as the action in our framework. This is a convenient choice, as parametric models are typically preferred for their tractability and interpretability, especially in lower-dimensional settings
\citep{gutierrez2005statistical,woody2021model}. 

Let ${\cal L}(\tilde{f},g_{\boldsymbol{\gamma}})$ be the loss function that quantifies the discrepancy between the summary $g_{\boldsymbol{\gamma}}$ and $\tilde{f}$. In our framework, the loss function is interpreted as the penalty incurred when selecting the summary $g_{\boldsymbol{\gamma}}$ through $\boldsymbol{\gamma}$ as an approximation for the model on the true density $f_{\boldsymbol{\theta}}$, under the lens of $\tilde{f}$.  Different loss functions can be employed for this objective, including the Hellinger and squared distances \citep{marin2005bayesian}. The choice of loss function is specified later in Subsection \ref{losschoice}.

Given the above definitions, selecting the optimal summary estimate for the reference model reduces to selecting the action \( \boldsymbol{\gamma} \) that minimizes the expected posterior predictive loss, as 
\begin{equation} \label{eq:min_lamb}
	\boldsymbol{\hat{\gamma}} := \argmin_{\boldsymbol{\gamma} \in \boldsymbol{\Gamma}} \mathbb{E}_{\boldsymbol{\tilde{y}}_{n} \mid \boldsymbol{y}} \left[ \mathcal{L}(\tilde{f}, g_{\boldsymbol{\gamma}}) \right] + \lambda P(\boldsymbol{\gamma}),
\end{equation}
where \( \boldsymbol{\tilde{y}}_{n} \) represents a random variable from \( \tilde{f} \), \( P(\boldsymbol{\gamma}) \) denotes a regularization term, and \( \lambda > 0 \) is the tuning parameter. From \eqref{eq:min_lamb}, \( \boldsymbol{\hat{\gamma}} \) is the optimal action with the corresponding density summary point estimate given by  \( \hat{g}_{\boldsymbol{\gamma}}(\cdot) := g(\cdot \mid \boldsymbol{\hat{\gamma}}) \).



In this paper, rather than using a penalty function in \eqref{eq:min_lamb}, we implement a greedy search over function spaces $\mathcal{G}_{\boldsymbol{\gamma}}^{k} := \{ g(\cdot | \boldsymbol{\gamma}^k); \boldsymbol{\gamma}^k \in\boldsymbol{\Gamma}^k, k \in \mathbb{Z}_{+} \}$ across \( k \in \{1, \dots, K_{\max}\} \). As a result, we are able to generate a sequence of optimal actions as
\begin{equation} \label{eq:min1}
	\boldsymbol{\hat{\gamma}}^k := \argmin_{\boldsymbol{\gamma}^k \in \boldsymbol{\Gamma}^k} \mathbb{E}_{\boldsymbol{\tilde{y}}_{n}\mid \boldsymbol{y}} \left[ {\cal L}(\tilde{f},g_{\boldsymbol{\gamma}})  \right],\quad k = 1,\dots,K_{\text{max}},
\end{equation}
which results in summary estimates \( \hat{g}_{\boldsymbol{\gamma}}^1, \dots, \hat{g}_{\boldsymbol{\gamma}}^{K_{\max}}  \), where $\hat{g}_{\boldsymbol{\gamma}}^k(\cdot) :=g(\cdot | \boldsymbol{\hat{\gamma}}^k)$,  with different levels of flexibility and where \( K_{\max} \) is chosen to be large enough to capture the complexity of the data. Here, it is important to note that $k$ is not necessarily the parametric dimension of the class of densities but a tuning parameter controlling model complexity. 

In relation to the process in \eqref{eq:min1}, a similar strategy was adopted in \cite{bolfarine2024decoupling} to generate summary estimates for factor loadings across different parametric dimensions. 




\subsection{Choice of the Loss Function and Class of Summaries}\label{losschoice}

\subsubsection{Negative Log-likelihood Loss Function}\label{loss}
Different loss functions can be used to measure the dissimilarity between the summary and the reference model $\tilde{f}$ \citep{marin2005bayesian}. In this paper, we focus on the negative log-likelihood function, defined as
\begin{equation} \label{eq:1}
	{\cal L}(\tilde{f},g_{\boldsymbol{\gamma}}) = -\log g_{\boldsymbol{\gamma}}^k(\boldsymbol{\tilde{y}}_n),
\end{equation}
which is a natural choice in this context due to its well-established optimal properties \citep{good1992rational}.  Moreover, under the decision-theoretic formulation in \eqref{eq:min1} and the loss function \eqref{eq:1}, selecting the optimal action is, up to an additive constant, equivalent to minimizing the KL divergence, defined as $\text{KL}(f \| g) = \int f \log \left( \frac{f}{g} \right) dy$, between the reference model $\tilde{f}$ and $g_{\boldsymbol{\gamma}}^k$. To demonstrate this, we express the expected posterior predictive loss in equation \eqref{eq:min1}  as
\begin{equation}\label{eq:KL}
	\mathbb{E}_{\boldsymbol{\tilde{y}}_{n}\mid \boldsymbol{y}} \left[ {\cal L}(\tilde{f},g_{\boldsymbol{\gamma}}) \right] = - \int \log g_{\boldsymbol{\gamma}}^k(\boldsymbol{\tilde{y}}_{n}) \tilde{f}(\boldsymbol{\tilde{y}}_{n}) d\boldsymbol{\tilde{y}}_{n}
	= \text{KL} ( \tilde{f}(\boldsymbol{\tilde{y}}_{n}) \| g_{\boldsymbol{\gamma}}^k(\boldsymbol{\tilde{y}}_{n})) + \text{H}(\tilde{f}(\boldsymbol{\tilde{y}}_{n})),
\end{equation}
where \( \text{H}(f) = - \int f \log f \, dy \) is the entropy of \( f \).


Notably, from \eqref{eq:KL}, we obtain the minimization of the forward KL divergence, which exhibits a mean-seeking behavior \citep{murphy2012machine}. Under this property, when \( \tilde{f} \) assigns a high probability to certain regions, the summary \( g_{\boldsymbol{\gamma}}^k \) must also assign a high probability to these regions. This ensures that \( g_{\boldsymbol{\gamma}}^k \) adequately captures modes and high-density regions of \( \tilde{f} \). This property is particularly important for our method, as these regions often correspond to modes or groups in the data, potentially indicating distinct clusters. 

The drawbacks of this choice and possible alternatives are discussed in Section~\ref{section7}.

\subsubsection{Finite Mixture Models Class of Summaries}

In principle, no restrictions are imposed on the class of the surrogate densities  $g_{\boldsymbol{\gamma}}^k$, provided that they possess the same support as $\boldsymbol{y}_i$. We focus on the case in which $\mathcal{G}_{\boldsymbol{\gamma}}^{k}$ is selected as the class of finite mixture models 
\citep{mclachlan2000finite, fruhwirth2006finite}. Thus, we update the class of functions \( \mathcal{G}_{\boldsymbol{\gamma}}^{k} \), defined by the actions given by \( \boldsymbol{\gamma}^k = ( \eta_1, \dots, \eta_k ;\boldsymbol{\gamma}_1, \dots, \boldsymbol{\gamma}_k) \), and where \( g(\cdot | \boldsymbol{\gamma}^k) \) is a mixture of densities with \( k \) components defined as
\begin{equation}\label{eq:finite}
	g(\cdot| \boldsymbol{\gamma}^k) = \sum_{q=1}^k \eta_q \, g(\cdot | \boldsymbol{\gamma}_q),
\end{equation}
where \( 0 \leq \eta_q \leq 1 \) for \( q = 1, \dots, k \), with the constraint \( \sum_{q=1}^k \eta_q = 1 \), and where \( g(\cdot| \boldsymbol{\gamma}_q) \) represents a kernel density defined by the summary parameters \( \boldsymbol{\gamma}_q \). In this paper, we adopt the Gaussian distribution as the kernel with $g(\cdot| \boldsymbol{\gamma}_q) = \frac{1}{\sqrt{2\pi\tau^2}}\exp(-(\cdot-\zeta)^2/2\tau^2)$, where $\boldsymbol{\gamma}_q = (\zeta,\tau^2)\in \mathbb{R}\times\mathbb{R}_{+}$, throughout the numerical applications in Section \ref{section6}. However, under our approach, other kernel choices are possible depending on the type of data and the desired level of approximation.

\subsection{Obtaining Finite Mixture Model Summary Estimates}\label{num_summary}

To obtain the optimal actions, we approximate the expected loss function in \eqref{eq:1}, under the class of functions \eqref{eq:finite} by applying Monte Carlo integration \citep{gamerman2006markov}. Specifically, given a sample \( \boldsymbol{\tilde{y}} = (\boldsymbol{\tilde{y}}_1, \dots, \boldsymbol{\tilde{y}}_{\tilde{N}})^T \) from \( \tilde{f} \) we approximate the expected negative log-likelihood loss with respect to the finite mixture model summary defined in \eqref{eq:finite}, as
\begin{equation}\label{montecarlo}
	\mathbb{E}_{\boldsymbol{\tilde{y}}_{n}\mid \boldsymbol{y}} \left[ {\cal L}(\tilde{f},g_{\boldsymbol{\gamma}})\right] = - \int \log g_{\boldsymbol{\gamma}}^k(\boldsymbol{\tilde{y}}_{n}) \tilde{f}(\boldsymbol{\tilde{y}}_{n}) d\boldsymbol{\tilde{y}}_{n}
	\approx -\frac{1}{\tilde{N}} \sum_{n=1}^{\tilde{N}} \log \sum_{q=1}^k \eta_q \, g(\boldsymbol{\tilde{y}}_n | \boldsymbol{\gamma}_q).
\end{equation}

It is straightforward to verify that minimizing \eqref{eq:min1} under the approximation in \eqref{montecarlo} is equivalent to maximizing the corresponding log-likelihood of the finite mixture model summary. Thus, for a fixed \( k \), we obtain the optimal set of actions \( \boldsymbol{\hat{\gamma}}^k \) as
\begin{equation} \label{optim}
	\boldsymbol{\hat{\gamma}}^k := \arg \max_{\boldsymbol{\gamma}^k \in \boldsymbol{\Gamma}^k}  \sum_{n=1}^{\tilde{N}} \log \sum_{q=1}^k \eta_q \, g(\boldsymbol{\tilde{y}}_n | \boldsymbol{\gamma}_q),
\end{equation}
resulting in the optimal summary point estimate \( \hat{g}^k_{\boldsymbol{\gamma}}\), which is a lower-dimensional finite mixture representation of the reference model $\tilde{f}$. In this paper, rather than focusing on the optimal action, we focus on the resulting density estimate $\hat{g}_{\boldsymbol{\gamma}}^k$.

As outlined in Section \ref{section21}, we apply a greedy search in \eqref{optim} across summaries with \( k = 1, \dots, K_{\max} \) components. This search produces a sequence of summary point estimates \( \hat{g}_{\boldsymbol{\gamma}}^1, \dots, \hat{g}_{\boldsymbol{\gamma}}^{K_{\max}} \), each providing a different level of flexibility. For this paper, we set \( K_{\max} \) to be of a similar magnitude to the number of components resulting from the original posterior. 

Notably, expression \eqref{optim} allows us to use several off-the-shelf procedures, such as the \texttt{mclust} R cran package from \cite{scrucca2023model}, among others, for obtaining \( \boldsymbol{\hat{\gamma}}^k \). The \texttt{mclust} package applies the EM algorithm \citep{dempster1977maximum} to solve problems as posted in \eqref{optim}. Our approach also draws a direct connection with the method presented in \cite{rodriguez2025martingale}, which applies the EM algorithm to obtain the posterior distribution for finite mixture model parameters under a Martingale Posterior approach \citep{fong2023martingale}. 

Next, we outline a framework for selecting the summary estimate \( \hat{g}_{\boldsymbol{\gamma}}^k \), through the optimal action $\boldsymbol{\hat{\gamma}}^k$, and the tuning parameter $k \in \{1, \dots, K_{\max}\}$, that best approximates \( f_{\boldsymbol{\theta}}\) through the posterior predictive distribution.

\subsection{Selecting the Lower-Dimensional Summary Estimate}\label{section23}


Under the choice of a finite mixture model summary in \eqref{eq:finite}, our main concern is the selection of the tuning parameter \( k\), since it controls the flexibility of the approximation over the reference model $\tilde{f}$. To achieve this, we propose a discrepancy function that captures the changes in fit between the summary estimates \( \hat{g}_{\boldsymbol{\gamma}}^1, \dots, \hat{g}_{\boldsymbol{\gamma}}^{K_{\max}} \) and $\tilde{f}$, as a function of $k$, as
\begin{equation} \label{eq:6}
	d_{n}^k(\tilde{f}, \hat{g}_{\boldsymbol{\gamma}}^k) = \log \frac{\hat{g}_{\boldsymbol{\gamma}}^k(\boldsymbol{\tilde{y}}_n)}{\tilde{f}(\boldsymbol{\tilde{y}}_n)}, \quad k = 1, \dots, K_{\max}, \quad n = 1,\dots, \tilde{N},
\end{equation}
where \( \boldsymbol{\tilde{y}} = (\boldsymbol{\tilde{y}}_1, \dots, \boldsymbol{\tilde{y}}_{\tilde{N}})^T \) is the same posterior predictive sample used for the optimization in \eqref{optim}. 

It is worth noting that, since the discrepancy function in \eqref{eq:6} depends on the posterior predictive distribution through the random variable \( \boldsymbol{\tilde{y}}_{n} \), we can obtain the expectation and uncertainty estimates of the quality of approximation of $\hat{g}_{\boldsymbol{\gamma}}^k$ to $\tilde{f}$ for every $k = 1,\dots,K_{\text{max}}$. Thus, for a fixed \( k \), the posterior predictive expectation of \eqref{eq:6} is given by
\begin{equation}\label{eqdiscr}
	\mathbb{E}_{\boldsymbol{\tilde{y}}_{n}\mid \boldsymbol{y}} \left[d_{n}^k\right] = -\int \tilde{f}(\boldsymbol{\tilde{y}}_n) \log \frac{\tilde{f}(\boldsymbol{\tilde{y}}_{n})}{\hat{g}_{\boldsymbol{\gamma}}^k(\boldsymbol{\tilde{y}}_{n})} \, d\boldsymbol{\tilde{y}}_{n}
	= -\text{KL}(\tilde{f} \parallel \hat{g}_{\boldsymbol{\gamma}}^k),
\end{equation}
where $d_{n}^k:=d_{n}^k(\tilde{f}, \hat{g}_{\boldsymbol{\gamma}}^k)$ and $\text{KL}(\tilde{f} \parallel \hat{g}_{\boldsymbol{\gamma}}^k)$ is the KL divergence between \( \hat{g}_{\boldsymbol{\gamma}}^k \) and \( \tilde{f}\).

From \eqref{eqdiscr}, we establish a heuristic for selecting the summary estimate \( \hat{g}_{\boldsymbol{\gamma}}^{k} \) for \( k = 1,  \dots, K_{\max} \) using the KL divergence as a reference. If the expected posterior predictive value of the discrepancy function is close to zero, \( \mathbb{E}_{\boldsymbol{\tilde{y}}_{n}\mid \boldsymbol{y}} \left[ d_{n}^k \right] \approx 0\), this suggests that the summary \( \hat{g}_{\boldsymbol{\gamma}}^{k} \) provides a good approximation of the model associated with \( \tilde{f} \). Otherwise, if \( \mathbb{E}_{\boldsymbol{\tilde{y}}_{n}\mid \boldsymbol{y}} \left[ d_{n}^k \right] < 0\), the summary estimate does not properly approximate $\tilde{f}$.


It is also important to note that the result in \eqref{eqdiscr} corresponds to the expected log-Bayes factor and the log-likelihood ratios associated with two distinct hypotheses \citep{kass1995bayes}. If the expectation 	$\mathbb{E}_{\boldsymbol{\tilde{y}}_{n}\mid \boldsymbol{y}} \left[d_{n}^k\right]$ is less than zero, there's evidence in favor of choosing $\tilde{f}$ as opposed to the summary estimate \( \hat{g}_{\boldsymbol{\gamma}}^{k} \) for future observations $\boldsymbol{\tilde{y}}_n$. If $\mathbb{E}_{\boldsymbol{\tilde{y}}_{n}\mid \boldsymbol{y}} \left[d_{n}^k\right]$ is close to zero, it indicates that the hypotheses are similar, and thus we have a favorable approximation of \( \hat{g}_{\boldsymbol{\gamma}}^{k} \) to the reference model.  We approximate the expectation using Monte Carlo integration, based on the posterior predictive sample  \( \boldsymbol{\tilde{y}} = (\boldsymbol{\tilde{y}}_1, \dots, \boldsymbol{\tilde{y}}_{\tilde{N}})^T \) as 
\begin{equation}\label{approx}
	\mathbb{E}_{\boldsymbol{\tilde{y}}_{n}\mid \boldsymbol{y}} \left[d_{n}^k \right] \approx \bar{d}^k(\tilde{f}, \hat{g}_{\boldsymbol{\gamma}}^k) = \frac{1}{\tilde{N}} \sum_{n=1}^{\tilde{N}} d_{n}^k(\tilde{f}, \hat{g}_{\boldsymbol{\gamma}}^k).
\end{equation}

We also assess the quality of the approximation of the summary estimate for the reference model by evaluating the uncertainty quantification, measured by the spread, i.e., the standard deviation of \eqref{eq:6}, $\text{sd}(d_{n}^k)$. We opted for this choice since the distribution of $d_{n}^k$ is multimodal, which in some cases might lead to difficulties in the interpretation, especially in terms of quantiles. If the spread is narrow and concentrated around $\mathbb{E}_{\boldsymbol{\tilde{y}}_{n}\mid \boldsymbol{y}} \left[d_{n}^k\right]$, there is a reduced uncertainty regarding the discrepancy between the estimate and the posterior predictive distribution. Conversely, if the spread is wide, there is greater uncertainty. 

We display the discrepancy function in \eqref{eq:6}, its respective expectation \eqref{approx}, and standard deviation $\text{sd}(d_{n}^k)$ with $n = 1,\dots,\tilde{N}$, and \( k = 1, \dots, K_{\max} \) in a plot. The resulting elbow plot visually represents the trade-off between model fit and the number of components, where each point $\mathbb{E}_{\boldsymbol{\tilde{y}}_{n}\mid \boldsymbol{y}} \left[d_{n}^k\right]$, reflects the degree of uncertainty in the approximation. To select $k\in \{1,\dots,K_{\text{max}}\}$ that results in the best summary approximation, we examine the plot for points closest to zero while ensuring that the corresponding interval remains narrow. Similar discrepancy plots were applied in different model summarization methods, including those by \cite{hanh-carvalho-2015}, \cite{woody2021model}, and \cite{bolfarine2024decoupling}.

Let \( K^* \) denote the tuning parameter selected under the aforementioned procedure, then the optimal summary point estimate is given by \(\hat{g}^{K^*}_{\boldsymbol{\gamma}}(\cdot) := g(\cdot| \boldsymbol{\hat{\gamma}}^{K^*})\), where \( \boldsymbol{\hat{\gamma}}^{K^*} \) is the optimal action defined by the space $\boldsymbol{\Gamma}^{K^*}$, that approximates the reference model. With finite mixture models as the class of summaries, this implies that \( \hat{g}^{K^*}_{\boldsymbol{\gamma}} \) is a finite mixture model density with \( K^* \) components.

\begin{algorithm}[t!]
	\caption{Summary estimates for densities}
	\label{algo_estimate}
	\begin{algorithmic}[1]
		\Require Posterior sample,
		$\boldsymbol{\tilde{y}}_n \sim \tilde{f}(\boldsymbol{\tilde{y}}),\;n = 1,\dots,\tilde N$.
		\Ensure Optimal actions
		$\boldsymbol{\hat\gamma}^1,\dots,\boldsymbol{\hat\gamma}^{K_{\max}}$
		and discrepancy functions
		$d_{n}^1,\dots,d_{n}^{K_{\max}}$.
		
		\State \textbf{Step 1: Generating optimal actions}
		\For{$k = 1$ to $K_{\max}$}
		\State
		\begin{equation*}
			{\boldsymbol{\hat{\gamma}}}^k
			= \argmax_{\boldsymbol{\gamma}^k\in\Gamma^k}
			\sum_{n = 1}^{\tilde N}
			\log \sum_{q = 1}^k \eta_q\,
			g(\boldsymbol{\tilde{y}}_n | \boldsymbol{\gamma}_{q}),
		\end{equation*}
		\EndFor
		\State \Return $\hat{\boldsymbol{\gamma}}^1,\dots,\hat{\boldsymbol{\gamma}}^{K_{\max}}$.
		
		\State\textbf{Step 2: Generating discrepancy functions}
		\For{$k = 1$ to $K_{\max}$}
		\For{$n = 1$ to $\tilde N$}
		\State
		\begin{equation}\label{algo_eq}
			\displaystyle
			d_{n}^k(\tilde f,\,\hat g_{\gamma}^k)
			= \log
			\frac{\hat g_{\gamma}^{k}(\boldsymbol{\tilde{y}}_n )}
			{\hat f(\boldsymbol{\tilde{y}}_n )},		
		\end{equation}
		\label{equation_algo}
		\EndFor
		\EndFor
		\State \Return $d_{n}^1,\dots,d_{n}^{K_{\max}}$.
	\end{algorithmic}
\end{algorithm}

Algorithm~\ref{algo_estimate} provides the procedure for obtaining the lower-dimensional summary estimate for densities. The summarization procedure begins by generating posterior predictive samples $\boldsymbol{\tilde{y}}_{n} \sim\tilde{f}(\boldsymbol{\tilde{y}})$, for $n = 1,\dots,\tilde{N}$. It is important to note that the sample from $\tilde{f}$ is not required to have the same size as the posterior sample. In Equation~\eqref{algo_eq} of Algorithm~\ref{algo_estimate}, the posterior predictive distribution $\tilde{f}$ is obtained over $\boldsymbol{\tilde{y}}_n$ as $ \hat{f}(\cdot) = \frac{1}{M}\sum_{m = 1}^{M}f(\cdot | \boldsymbol{\theta}^{(m)})$, where $\boldsymbol{\theta}^{(m)}$ represents the samples drawn from the posterior distribution of the reference model, $p(\boldsymbol{\theta}|\boldsymbol{y})$. 

In Section \ref{section6}, we demonstrate the effectiveness of the discrepancy plots across various numerical applications.

\section{Posterior Summarization for Densities}\label{section3}

In previous summarization methods, such as those proposed by \cite{hanh-carvalho-2015, puelz2017variable, bashir2019post, bolfarine2024decoupling}, the summary estimate was the final objective. In this section, we extend our analysis by projecting the original posterior distribution onto a lower-dimensional posterior summary. This concept was first introduced by \cite{woody2021model} to summarize the posterior distribution of Gaussian process regression in terms of linear and additive models. This approach enables summary uncertainty quantification through credible intervals, which was not feasible with the summary point estimates obtained in the previous section.


To generate the summary posterior, we project the original posterior distribution onto the same class of functions defined by the optimal estimate. Essentially, we build the posterior distribution around \( \hat{g}^{K^*}_{\boldsymbol{\gamma}} \) based on the original posterior \(p(\boldsymbol{\theta} | \boldsymbol{y}) \), resulting in uncertainty estimates for the lower-dimensional summary. We obtain the projection by minimizing the loss between $\tilde{f}$ and the summary density $g_{\boldsymbol{\gamma}}^{K^*} \in {\cal G}^{K^*}$, using the same loss function and class of density summaries as in the process from Section \ref{section21}. The posterior projection is defined by approximating \( \boldsymbol{\gamma}^{K^*}\) in \( g_{\boldsymbol{\gamma}}^{K^*} \) so that it matches \( \tilde{f} \) over the original posterior \( p(\boldsymbol{\theta} | \boldsymbol{y}) \), by solving
\begin{equation}\label{possum}
	\boldsymbol{\gamma}' := \argmin_{\boldsymbol{\gamma} \in \boldsymbol{\Gamma}^{K^*}} \mathcal{L}(\tilde{f}, g_{\boldsymbol{\gamma}}^{K^*}),
\end{equation}  
resulting in the lower-dimensional posterior summary, which we represent as  \( \boldsymbol{\gamma}' := p(\boldsymbol{\gamma}^{K^*} | \boldsymbol{y}) \), from which we obtain the posterior density summary \( g(\cdot|\boldsymbol{\gamma}') \).




Here, we highlight the distinction between the summary point estimate in \eqref{eq:min1} and  \( \boldsymbol{\gamma}' \), which is the lower-dimensional characterization of the posterior distribution. These quantities are analogous to the Bayes estimator and the posterior distribution of a parameter. Moreover, as noted by \cite{woody2021model}, the estimates obtained in \eqref{eq:min1} and the lower posterior summary \eqref{possum} are generally not equivalent
\begin{equation*}
	\argmin_{\boldsymbol{\gamma} \in \boldsymbol{\Gamma}} \mathbb{E}_{\boldsymbol{\tilde{y}}_{n} \mid \boldsymbol{y}} \left[ \mathcal{L}(\tilde{f}, g_{\boldsymbol{\gamma}}) \right] \neq  \mathbb{E}_{\boldsymbol{\tilde{y}}_{n} \mid \boldsymbol{y}} \left[\argmin_{\boldsymbol{\gamma} \in \boldsymbol{\Gamma}} \mathcal{L}(\tilde{f}, g_{\boldsymbol{\gamma}}) \right].
\end{equation*}

It is important to note that minimizing \eqref{possum} presents significant challenges, particularly due to the choice of the loss function \eqref{eq:1} and the finite mixture model class of summaries \eqref{eq:finite} since these choices do not yield a closed-form solution. 

To address this issue, we approximate the solution using Monte Carlo simulation. Let \( \boldsymbol{\theta}^{(1)}, \dots, \boldsymbol{\theta}^{(M)} \) be a sample of size \( M \) drawn from the original posterior \( p(\boldsymbol{\theta} | \boldsymbol{y}) \). For each \( \boldsymbol{\theta}^{(m)} \), we generate posterior predictive sample of size $H$, \( \boldsymbol{\tilde{y}}_1^{(m)}, \dots, \boldsymbol{\tilde{y}}_H^{(m)} \) from the conditional density \( f(\boldsymbol{\tilde{y}}_h^{(m)} | \boldsymbol{\theta}^{(m)}) \).  From which we obtain the lower-dimensional posterior summary as
\begin{equation}\label{possum_mc}
	\boldsymbol{\gamma}'^{(m)}:= \argmax_{\boldsymbol{\gamma}\in \boldsymbol{\Gamma}^{K^*}}\sum_{h = 1}^H\log\sum_{q = 1}^{K^*}\boldsymbol{\eta}_q  g(\boldsymbol{\tilde{y}}_{h}^{(m)}|\boldsymbol{\gamma}_q^{K^*}), \quad m = 1,2,\dots,M,
\end{equation}
producing a sequence $\boldsymbol{\gamma}'^{(1)},\dots,\boldsymbol{\gamma}'^{(M)}$ which are samples from $\boldsymbol{\gamma}'$.  Algorithm \ref{algo2} provides a more detailed overview of the procedure.

As with the optimization in \eqref{optim}, we can solve expression \eqref{possum_mc} using various off-the-shelf procedures, depending on the choice of the kernel for the summary. We adopt the \texttt{mclust} package from CRAN \citep{scrucca2023model} to obtain \( \boldsymbol{\gamma'}^{(m)} \) for each \( m = 1,\dots,M \).

In this paper, we focus on the posterior expectation and uncertainty quantification for the posterior summary density \( g(\cdot|\boldsymbol{\gamma}') \), induced by \( \boldsymbol{\gamma}' \), rather than the posterior summary $\boldsymbol{\gamma}'$. To quantify the uncertainty for \( g(\cdot|\boldsymbol{\gamma}') \), we construct 95\% credible intervals for the density function at each point \( \boldsymbol{y}_i \). Specifically, we compute the 2.5th and 97.5th percentiles, $\left[g(\cdot|\boldsymbol{\gamma'})_{2.5}, g(\cdot|\boldsymbol{\gamma}')_{97.5}\right]$, based on the of the posterior sample \( g(\cdot|\boldsymbol{\gamma'}^{(m)}) \) for \( m = 1, \dots, M \).  We obtain the posterior expected estimate \( g(\cdot|\boldsymbol{\gamma}') \) as 
\begin{equation}\label{eqpost}
	\bar{g}^{K^*}(\cdot|\boldsymbol{\gamma}') = \frac{1}{M} \sum_{m=1}^{M} g(\cdot|\boldsymbol{\gamma'}^{(m)}).
\end{equation}

The immediate result of the lower-dimensional posterior summarization approach is presented in Section~\ref{section1} in Figure \ref{BP_galaxy_all} plots (a) and (b). In both cases a Gaussian mixture model summary is shown for the random Bernstein polynomial model with three, $K^* = 3$, and four, $K^* = 4$, components, respectively. The gray ribbon represents the corresponding 95\% credible interval for the summary, $\left[g(\cdot|\boldsymbol{\gamma'})_{2.5}, g(\cdot|\boldsymbol{\gamma'})_{97.5}\right]$, and the expected posterior of the summary, $\bar{g}^{K^*}_{\boldsymbol{\gamma}}:=\bar{g}^{K^*}(\cdot|\boldsymbol{\gamma}')$, is depicted as a red line. 


\begin{algorithm}[t]
	\caption{Posterior Summarization for densities}
	\label{algo2}
	\begin{algorithmic}[1]
		\Require Posterior samples $\boldsymbol{\theta}^{(m)}\sim p(\boldsymbol{\theta}| \boldsymbol{y}), m=1,\dots,M$.
		\Ensure Samples $\boldsymbol{\gamma}'^{(1)},\dots,\boldsymbol{\gamma}'^{(M)}$.
		
		\State \textbf{Step 1: Generate posterior predictive samples}
		\For{$m = 1$ to $M$}
		\For{$h = 1$ to $H$}
		\State $\displaystyle
		\boldsymbol{\tilde y}_{h}^{(m)}
		\sim f\bigl(\boldsymbol{\tilde y}_{h}| \boldsymbol{\theta}^{(m)}\bigr),
		$
		\EndFor
		\EndFor

		\Return $(\;\boldsymbol{\tilde y}_{1}^{(m)},\dots,\boldsymbol{\tilde y}_{H}^{(m)}),m=1,\dots,M$.
		
		\State \textbf{Step 2: Compute posterior summary}
		\For{$m = 1$ to $M$}
		\State 
		\begin{equation*}
			\displaystyle
			\boldsymbol{\gamma}'^{(m)}
			= \argmax_{\boldsymbol{\gamma}\in \boldsymbol{\Gamma}^{K^*}}
			\sum_{h = 1}^H
			\log
			\sum_{q = 1}^{K^*} \eta_q \,
			g\bigl(\boldsymbol{\tilde y}_{h}^{(m)}| \boldsymbol{\gamma}_q^{K^*}\bigr),
		\end{equation*}
		\EndFor
		\Return $\;\boldsymbol{\gamma}'^{(1)},\dots,\boldsymbol{\gamma}'^{(M)}$.
	\end{algorithmic}
\end{algorithm}

\section{Posterior Cluster Summaries}\label{section5}

An additional contribution of this paper is the extension of the posterior summarization method presented in Sections \ref{section2} and \ref{section3} to derive cluster summary point estimates accompanied by uncertainty quantification. The proposed procedure aims to preserve the predictive accuracy of the original model while producing parsimonious and interpretable cluster summaries, even in settings where membership allocation was not originally intended.

One of the main advantages of the proposed approach is its modularity, since within a single reference model, it enables the derivation of distinct cluster summaries with different data allocation behaviors. In this case, the modularity is determined by the choice of the loss function. In this paper, we examine two alternatives; the first is the log-score loss, linked to the finite mixture model summaries introduced in Section \ref{losschoice}. The second is the k-means loss, related to the k-means clustering method \citep{MacQueen1967}.

Under both loss functions, we define the cluster summary to consist of $K^*$ groups. This value corresponds to the number of components in the finite mixture model density summary point estimate, identified through the heuristic outlined in Section \ref{section23}. The motivation for this specification is discussed in Section \ref{losschoice}.

\subsection{Conditional Probability Allocation Summaries} \label{section51}


As is well known, finite mixture models as defined in Equation \eqref{equation1}, Section \ref{section1}, can also be expressed using latent variables $z_i = q$, with $q \in \{1,\dots,K\}$, which indicate the component allocation for each observation. The model can then be rewritten as $\boldsymbol{y}_i | z_i \sim f(\boldsymbol{y}_i | \boldsymbol{\theta}_{z_i})$, with conditional allocation probability $\omega_i(q) := \mathbb{P}(z_i = q)$, given by
\begin{equation}\label{eq_prob_cond}
	\omega_i(q) = \frac{\omega_{q}\, f(\boldsymbol{y}_i | \boldsymbol{\theta}_{q})}{\sum_{l=1}^{K} \omega_l \, f(\boldsymbol{y}_i | \boldsymbol{\theta}_l)},\quad q\in \left\{1,\dots,K\right\},\quad i = 1,\dots,N,
\end{equation}
which highlights the utility of finite mixture models for clustering \citep[see][]{fruhwirth2006finite,mclachlan2019finite}. 

Under our approach, both the conditional probability and the corresponding component allocation are treated as actions within a decision-theoretic framework. We next demonstrate how to obtain these actions, followed by the procedure for deriving lower-dimensional posterior summaries to quantify uncertainty in the allocations.



Let $\hat{g}_{\gamma}^{K^*}$ denote the finite mixture model density summary estimate obtained under the method outlined in Section \ref{section3}, defined by the actions $
\boldsymbol{\hat{\gamma}}^{K^*} = (\hat{\eta}_1, \dots, \hat{\eta}_{K^*}; \boldsymbol{\hat{\gamma}}_1, \dots, \boldsymbol{\hat{\gamma}}_{K^*})$. In our framework, \eqref{eq_prob_cond} specifies the probability of a latent categorical variable $c_n \in \{1,\dots,K^*\}$ in the surrogate model $\cdot | c_n \sim g(\cdot |\boldsymbol{\hat{\gamma}}_{c_n})$. In this case, the conditional probability summary membership for the $i$-th original observation $\boldsymbol{y}_i$, denoted by $\hat{\eta}_i(q) := \mathbb{P}(c_i = q)$, is obtained as
\begin{equation}\label{cond_prob}
	\hat{\eta}_i(q) = \frac{\hat{\eta}_{q}\, g(\boldsymbol{y}_i | \boldsymbol{\hat{\gamma}}_{q})}{\sum_{l=1}^{K^*} \hat{\eta}_l g(\boldsymbol{y}_i| \boldsymbol{\hat{\gamma}}_l)},\quad q\in \left\{1,\dots,K^*\right\}.
\end{equation}

From the above conditional probability summary, we derive the summary estimate for the cluster membership for $\boldsymbol{y}_i$ as
\begin{equation}\label{clust_alloc}
	\hat{c}_i = \argmax_{q \in \{1,\dots,K^*\}}\hat{\eta}_i(q),\quad q\in \left\{1,\dots,K^*\right\},\quad i = 1,\dots,N.
\end{equation}


Although useful for summarizing membership allocation, the conditional probability estimates in \eqref{cond_prob} and the cluster allocation estimates in \eqref{clust_alloc} are akin to Bayes point estimates and therefore are not equipped with uncertainty quantification. To address this limitation, we follow the procedure introduced in Section \ref{section3}, and project the full posterior distribution onto the conditional probability function. This is achieved via the lower-dimensional posterior summary $\boldsymbol{\gamma}'$ obtained in \eqref{possum_mc}, with samples $\boldsymbol{\gamma'}^{(m)} = ( \eta_{1}^{(m)}, \dots, \eta_{K^*}^{(m)} ;\boldsymbol{\gamma'}_{1}^{(m)}, \dots, \boldsymbol{\gamma'}_{K^*}^{(m)})$, for $m=1,\dots,M$, which is then projected onto the conditional probability posterior summary \eqref{cond_prob} for the $i$-th observation $\boldsymbol{y}_i$ resulting in
\begin{equation}\label{cond_possum}
	\eta_{i}^{(m)}(q)  = \frac{\eta^{(m)}_{q}\, g(\boldsymbol{y}_i | \boldsymbol{\gamma'}_{q}^{(m)})}{\sum_{l=1}^{K^*} \eta_l^{(m)} \, g(\boldsymbol{y}_i | \boldsymbol{\gamma'}^{(m)}_l)},\quad q\in \left\{1,\dots,K^*\right\}.
\end{equation}

From \eqref{cond_possum}, we can then compute the posterior projected cluster allocation for each $m = 1,\dots,M$ posterior draw as
\begin{equation}\label{cluster_kmax}
c^{(m)}_i = \argmax_{q \in \{1,\dots,K^*\}}\eta_{i}^{(m)}(q),\quad i = 1,\dots,N,
\end{equation}
resulting in a projected posterior distribution of cluster assignments for each observation. From \eqref{cond_possum} and \eqref{cluster_kmax}, we can easily obtain credible intervals for the conditional probability and cluster allocation summaries. In this paper, we quantify uncertainty in the allocation summaries by analyzing the expected assignment variation from \eqref{cluster_kmax} across posterior draws as
\begin{equation}\label{Uncert_cond}
	\text{cond-Uncert}_{i} = 1 - \max_{q \in \{1,\dots,K^*\}}\frac{1}{M}\sum_{m = 1}^M \mathbb{I}\{c_{i}^{(m)}=q\},
\end{equation}
where $\mathbb{I}\{\cdot\}$ is the indicator function. \cite{Bouveyron_Celeux_Murphy_Raftery_2019} use a similar uncertainty quantification measure for cluster allocation.


\subsection{k-means  Summary and Uncertainty Quantification} \label{section42}




The standard k-means method partitions data into discrete groups by assigning each observation to the nearest cluster center, or centroid \citep{bishop2006pattern}. This assignment is achieved by minimizing the squared, or Euclidean, distance between data points and their respective centroids. Cluster memberships and centroids can be efficiently determined using the k-means algorithm \citep{MacQueen1967}.

Under our approach, we employ the same posterior predictive sample, $\boldsymbol{\tilde{y}} = (\boldsymbol{\tilde{y}}_1,\dots,\boldsymbol{\tilde{y}}_{\tilde{N}})^T$, used to determine $K^*$, to obtain the optimal cluster allocations and centroids, which are treated as actions in our decision-theoretic framework. The k-means loss function is defined as
\begin{equation}\label{kmeansloss}
	{\cal L}(\boldsymbol{\xi},\tilde{f},c_{n}) = \sum_{n = 1}^{\tilde{N}}\sum_{q = 1}^{K^*}\mathbb{I}\{c_n = q\}\|\boldsymbol{\tilde{y}}_n-\boldsymbol{\xi}_q\|_{2}^2,
\end{equation}
where $\boldsymbol{\xi} = \{\boldsymbol{\xi}_1,\dots,\boldsymbol{\xi}_{K^*}\}$ are the centroids, and $\|\boldsymbol{y}\|_{2}^2 = (\sqrt{\boldsymbol{y}^T\boldsymbol{y}})^2$ denotes the squared Euclidean distance \citep{hastie2009}. The expected sum of squared distances between the predictive samples and the centroids in \eqref{kmeansloss} is minimized to obtain the optimal summary estimates for both centroids and cluster allocations, as
\begin{equation}\label{min}
	(\boldsymbol{\hat{\xi}},\hat{c}_n):= \argmin_{\substack{\boldsymbol{\xi}_q\in \mathbb{R} \\ c_n\in\{1,\dots,K^*\}}} \mathbb{E}_{\boldsymbol{\tilde{y}}_n|\boldsymbol{y}}\left[\sum_{n = 1}^{\tilde{N}}\sum_{q = 1}^{K^*}\mathbb{I}\{c_n = q\}\|\boldsymbol{\tilde{y}}_n-\boldsymbol{\xi}_q\|_{2}^2\right].
\end{equation}

We approximate \eqref{min} using Monte Carlo integration, resulting in the minimization
\begin{equation}\label{kmeans_min}
	(\boldsymbol{\hat{\xi}},\hat{c}_n):= \argmin_{\substack{\boldsymbol{\xi}_q\in \mathbb{R} \\ \boldsymbol{c}_n\in\{1,\dots,K^*\}}} \frac{1}{\tilde{N}} \sum_{n = 1}^{\tilde{N}}\sum_{q = 1}^{K^*}\mathbb{I}\{c_n = q\}\|\boldsymbol{\tilde{y}}_n-\boldsymbol{\xi}_q\|_{2}^2,
\end{equation}
which can be easily solved using the standard k-means algorithm of \cite{MacQueen1967}. The resulting centroids, $\boldsymbol{\hat{\xi}} =\{\boldsymbol{\hat{\xi}}_{1},\dots,\boldsymbol{\hat{\xi}}_{K^*}\}$ from \eqref{kmeans_min}, are then used to assign the original observations $\boldsymbol{y}_i$ to the nearest centroid as a prediction, yielding the cluster allocation summaries
\begin{equation}\label{kmeans_aloc}
	\hat{c}_i =\argmin_{q\in\{1,\dots,K^*\}}\|\boldsymbol{y}_i- \boldsymbol{\hat{\xi}}_{q}\|_{2}^2, \quad i = 1,\dots N.
\end{equation}


As is well known, the standard k-means clustering method does not provide uncertainty quantification for membership allocation. Additionally, as noted in Section \ref{section51}, the summaries in \eqref{kmeans_aloc} are comparable to a Bayes point estimate and thus does not provide uncertainty quantification.

We again apply the posterior summarization approach described in Section \ref{section3}, to project the posterior distribution and obtain summaries that capture uncertainty in cluster allocation under the k-means loss. We project the posterior distribution through posterior predictive samples \( \boldsymbol{\tilde{y}}_1^{(m)}, \dots, \boldsymbol{\tilde{y}}_H^{(m)} \) from \( f(\boldsymbol{\tilde{y}}_h^{(m)}| \boldsymbol{\theta}^{(m)}) \)  in the k-means loss defined in \eqref{kmeansloss} as 
\begin{equation}\label{kmeans_possum2}
	(\boldsymbol{\xi}^{(m)},c^{(m)}_h):=  \argmin_{\substack{\boldsymbol{\xi}_q\in \mathbb{R} \\ c_n\in\{1,\dots,K^*\}}} \sum_{h = 1}^{H}\sum_{q = 1}^{K^*}\mathbb{I}\{c^{(m)}_h = q\}\|\boldsymbol{\tilde{y}}_{h}^{(m)}-\boldsymbol{\xi}_q\|_{2}^2,
\end{equation}
which results in the posterior predictive projection over the centroids $\boldsymbol{\xi}^{(m)}=\{\boldsymbol{\xi}_{1}^{(m)},\dots,\boldsymbol{\xi}^{(m)}_{K^*}\}$, for  \( m = 1,\dots,M\). For each posterior observation, the summary estimate of cluster allocation is then given by
\begin{equation}\label{kmeans_possum}
	c_{i}^{(m)} =\argmin_{q\in\{1,\dots,K^*\}}\|\boldsymbol{y}_i - \boldsymbol{\xi}^{(m)}_q\|_{2}^2, \quad i = 1,\dots,N,\quad m = 1,\dots,M,
\end{equation}
which enables uncertainty quantification in the assignments for the original observations. Similar to the uncertainty quantification procedure adopted for conditional probability allocation discussed in Section \ref{section51}, we derive uncertainty measures from \eqref{kmeans_possum} by analyzing the expected variation in cluster assignments across posterior draws, as
\begin{equation}\label{Uncert_kmeans}
	\text{k-Uncert}_{i} = 1 - \max_{q \in \{1,\dots,K^*\}}\frac{1}{M}\sum_{m = 1}^M \mathbb{I}\{c_{i}^{(m)}=q\}.
\end{equation}

Figure~\ref{BP_galaxy_all}, plots (a) and (b) in Section~\ref{section1}, illustrates an application of posterior summary estimates for cluster allocation under the conditional and k-means losses, respectively. The cluster allocation estimates and associated uncertainties were generated using three and four components. 



A drawback of the approaches presented in this section, observed in the numerical experiments, is label switching  \citep{papastamoulis2010artificial,fruhwirth2011dealing,malsiner2016model}, arising from the numerical methods used to obtain the cluster allocations and under the conditional allocation and k-means loss, respectively.
To  address this, we order the location parameters in \eqref{cond_possum} as $\boldsymbol{\gamma'}_{1}^{(m)}<\dots < \boldsymbol{\gamma'}_{K^*}^{(m)}$ and the centroids in \eqref{kmeans_possum2} as $\boldsymbol{\xi}_{1}^{(m)}<\dots<\boldsymbol{\xi}_{K^*	}^{(m)}$, in a similar process to \cite{fruhwirth2011dealing} for handling label switching in posterior samples of finite mixture models.

\section{Numerical Applications}\label{section6}

Our goal in this section is to demonstrate the flexibility of our method in generating density and cluster allocation summaries under different models. The results are provided in \url{https://github.com/hbolfarine/mix-possum}, which includes scripts that reproduce the analyses for both simulated and real data.

\subsection{Univariate Models}\label{unisim}

For univariate data, we applied our posterior summarization method to the following models: the random Bernstein polynomial \citep{petrone1999random}, the Dirichlet process mixture \citep[DPM;][]{ferguson1983bayesian, escobar1995bayesian}, and mixture of finite mixtures \citep[MFM;][]{richardson1997bayesian,miller2018mixture}. A brief description of these methods, the hyperparameters, and the posterior sampling conditions is given in Supplementary Material \ref{secA1}.


\subsubsection{Simulated Data}\label{simdata}

In the first simulation scenario, similar to \cite{papastamoulis2010artificial}, we generated $N = 600$ observations from a five-component Gaussian mixture model, in which two of the components overlap in their location and mixing weights. The data-generating process is specified as  $\boldsymbol{y}_i \sim \sum_{q=1}^5 \omega_q , N(\boldsymbol{y}_i | \mu_q, \sigma_q^2),$ where $\mu = (19, 19, 23, 20, 33)$ denotes the locations, $\sigma^2 = (5, 1, 1, 1, 2)$ the variances, and $\omega = (0.2, 0.2, 0.25, 0.2, 0.15)$ the mixing weights.

\begin{figure*}[t!]
	\centering
	\includegraphics[width=1\textwidth]{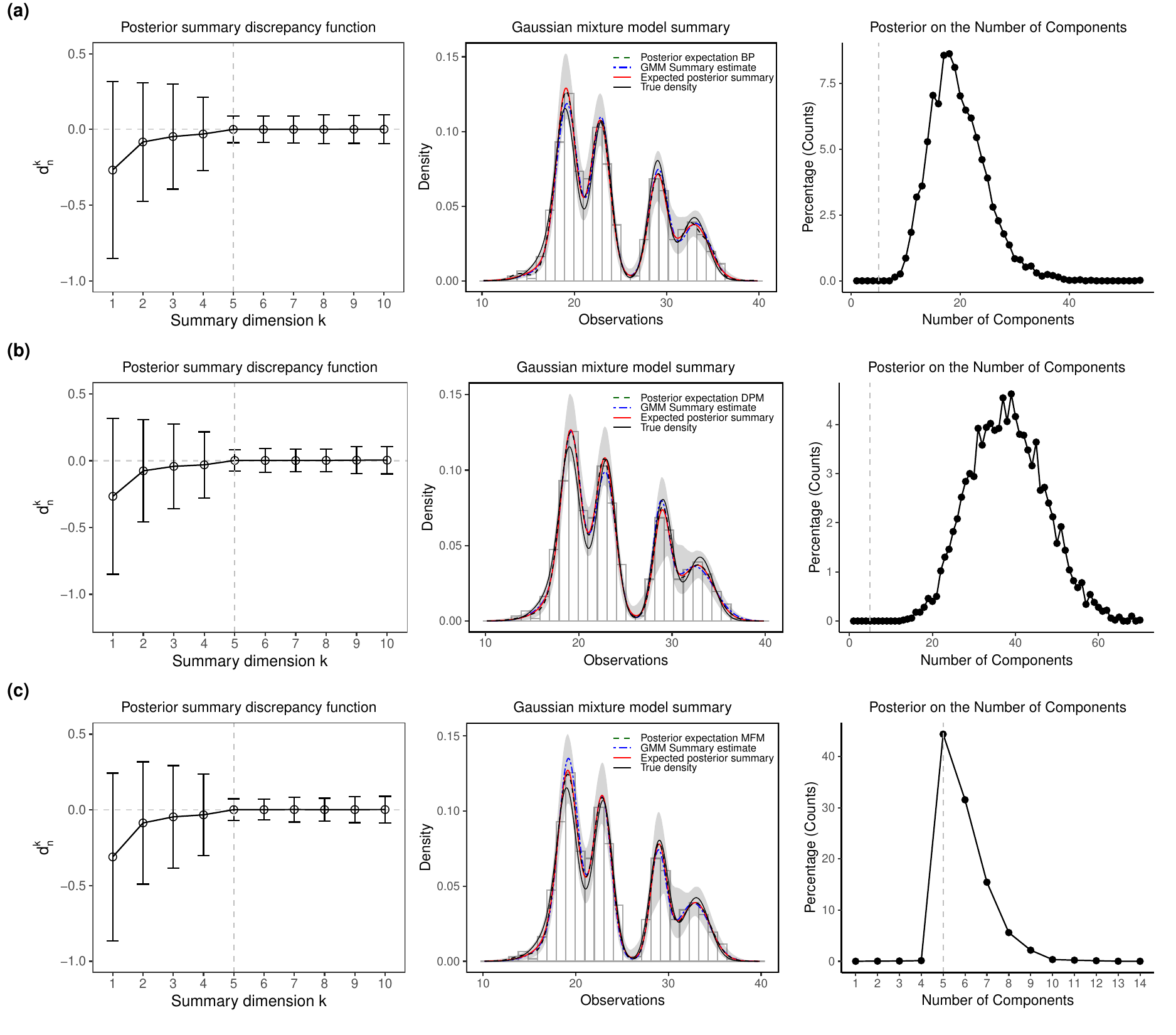}
	\caption{Posterior Summaries for the simulated data: In Figures (a), (b), and (c), the plots on the left display the discrepancy function, $d_{n}^k$, under the respective priors: BP, DPM, and MFM. In the center plots, the black dashed line represents the expected posterior density estimated for the different models. The blue line is the Gaussian summary estimate, $\hat{g}_{\boldsymbol{\gamma}}^{K^*}$, with $K^* = 5$ components. Additionally, we display the 95\% credible ribbon of the posterior summary (in gray) and the expected average, $\bar{g}_{\boldsymbol{\gamma}}^{K^*}$, (red line), for the $K^* = 5$ component-Gaussian mixture model posterior summaries. The right plots show the posterior probabilities for the number of components under the BP, DPM, and MFM priors, respectively, the dashed line represents the true number of components.}\label{sim_examp_all}
\end{figure*}

We first obtained the posterior distributions from the BP, DPM, and MFM models, from which we generated samples of size $\tilde{N}=2000$ from the respective posterior predictive distributions $\tilde{f}$. Following Section \ref{num_summary}, we applied Algorithm \ref{algo_estimate} and obtained the sequence of finite mixture model summary estimates $\hat{g}_{\boldsymbol{\gamma}}^k$, with Gaussian kernels and a maximum of $K_{\text{max}}=10$ components. From these estimates, we derived the discrepancy function plots $d_{n}^k$ for $n = 1,\dots,2000$ and $k = 1,\dots,10$, as defined in \eqref{eq:6}. Figure~\ref{sim_examp_all} displays the average discrepancy function, $\bar{d}_{n}^k$ (points), as defined in \eqref{approx} alongside plus or minus one standard deviation, $\text{sd}(d_{n}^k)$ (bars). The discrepancy plots are shown on the left panels of (a), (b), and (c), generated under the BP, DPM, and MFM models, respectively.

Using the heuristic presented in Section \ref{section23}, in the discrepancy plots, we identified that the summary $g_{\boldsymbol{\gamma}}^k$ with $k = 5$ components provided a good approximation for the predictive posterior under all models. This is supported by the average discrepancy $\bar{d}_n^k$ being near zero. Moreover, the uncertainty for summaries with $k = 5$ components, measured by $\text{sd}(d_{n}^k)$, is small, indicated by the narrow interval. For summaries with $k<5$, the uncertainty increases substantially, reflecting a poor approximation of summaries in these cases. Following Section \ref{section3}, we projected the posterior of the models via Algorithm \ref{algo2}, in a summary with $K^* = 5$ components, and obtained the lower-dimensional parameter summary $\boldsymbol{\gamma}'$ and the corresponding density $g(\cdot|\boldsymbol{\gamma}')$, which includes uncertainty quantification. 

The center panels of Figure \ref{sim_examp_all} ((a), (b), and (c)) show the simulated data alongside the expected posterior under the BP, DPM, and MFM priors (black dashed line). The blue line represents the Gaussian summary estimate, $\hat{g}_{\boldsymbol{\gamma}}^{K^*}$, with $K^* = 5$ components. The gray ribbons indicate its 95\% credible interval of  $g(\cdot|\boldsymbol{\gamma}')$. The red line depicts the posterior average of the surrogate density, $	\bar{g}^{K^*}_{\boldsymbol{\gamma}}$, defined in \eqref{eqpost}. We can observe that in all cases, the true density and the posterior averages fall within the bounds of their respective summaries. The panels on the right display the posterior distribution of the number of components for each model.


\subsubsection{Posterior Summary Estimate Approximation}\label{section512}

In the second simulation scenario,  we evaluated how well the summary density estimate $\hat{g}_{\boldsymbol{\gamma}}^k$, and the average posterior summary projection $\bar{g}_{\boldsymbol{\gamma}}^k$, approximated the true underlying distribution $f_{\boldsymbol{\theta}}$ under the DPM model as the sample size increased. We generated 100 independent datasets of sample sizes of $N\in \{100,250, 1000\}$, under the same data-generating process as in the simulation study in Section \ref{simdata}. We maintained the same hyperprior specification and posterior sampling conditions. For the summary estimate, we generated $\tilde{N} = 2000$ posterior predictive samples. We then followed Algorithms \ref{algo_estimate} and \ref{algo2} to obtain the lower-dimensional posterior approximation estimates $\hat{g}_{\boldsymbol{\gamma}}^{K^*}$ and $\bar{g}_{\boldsymbol{\gamma}}^{K^*}$. In this study, we set $K^* = 5$, as we were interested in evaluating the performance of the different summaries in terms of density approximation as the data size increased for the best summary approximation given the discrepancy plot in Figure \ref{sim_examp_all}, plot (b), left panel. We used the Hellinger distance defined as 
\begin{equation}\label{hellinger}
	H(f_{\boldsymbol{\theta}}, g_{\boldsymbol{\gamma}}) = \frac{1}{\sqrt{2}} \left( \int \left( \sqrt{f_{\boldsymbol{\theta}}(\boldsymbol{y}_i)} - \sqrt{g_{\boldsymbol{\gamma}}(\boldsymbol{y}_i)} \right)^2 \, d\boldsymbol{y}_i \right)^{1/2},
\end{equation}
to evaluate the approximation of the summaries densities, $g_{\boldsymbol{\gamma}} \in \{\hat{g}_{\boldsymbol{\gamma}}^{K^*},\bar{g}_{\boldsymbol{\gamma}}^{K^*}\}$  to the true density $f_{\boldsymbol{\theta}}$. We approximated the integral in \eqref{hellinger}, as the average of the samples $\boldsymbol{y}_i$. We compared the performance of these summaries to the expected posterior density of the DPM model.

\begin{figure}[t!]
	\centering
	\includegraphics[width=.6\linewidth]{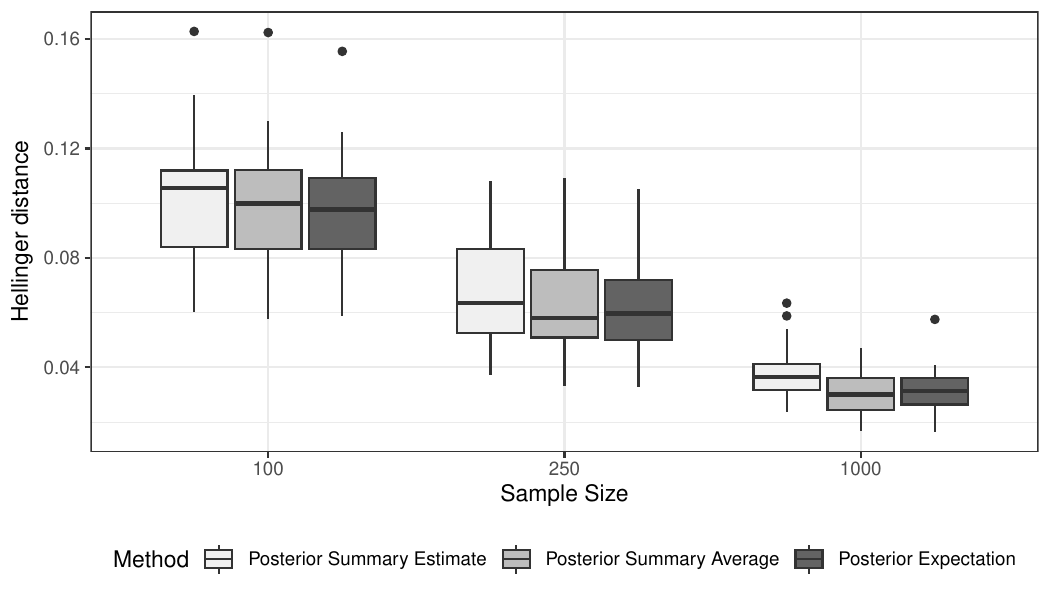}
	\caption[]{Hellinger distance between the true density and both the summary estimate, $\hat{g}_{\boldsymbol{\gamma}}^{K^*}$, and the average posterior summary, $\bar{g}_{\boldsymbol{\gamma}}^{K^*}$, with $K^* = 5$. For each sample size $N \in \{100, 250, 1000\}$, 100 independent data sets were generated. We compare the performance of these summaries to the expected posterior density.}
	\label{DPM_hellinger}
\end{figure}

Figure \ref{DPM_hellinger} we observe that the Hellinger distance in \eqref{hellinger}, between the density summaries and the true density is going to zero, and further, that the rate of convergence appears to be nearly identical for the $\hat{g}_{\boldsymbol{\gamma}}^{K^*}$ and the $\bar{g}_{\boldsymbol{\gamma}}^{K^*}$, and similar to the expected posterior density.

\subsubsection{Galaxy Data}\label{section613}

%


We further demonstrate the performance of our method using the galaxy data, a well-known benchmark dataset \citep{fruhwirth2019handbook}. This dataset has been extensively analyzed under a variety of parametric and nonparametric models, including those by \cite{escobar1995bayesian}, and \cite{richardson1997bayesian}. In most analyses the number of groups in the data is considered to be between three and seven.

\begin{figure*}[t!]
	\centering
	\includegraphics[width=1\textwidth]{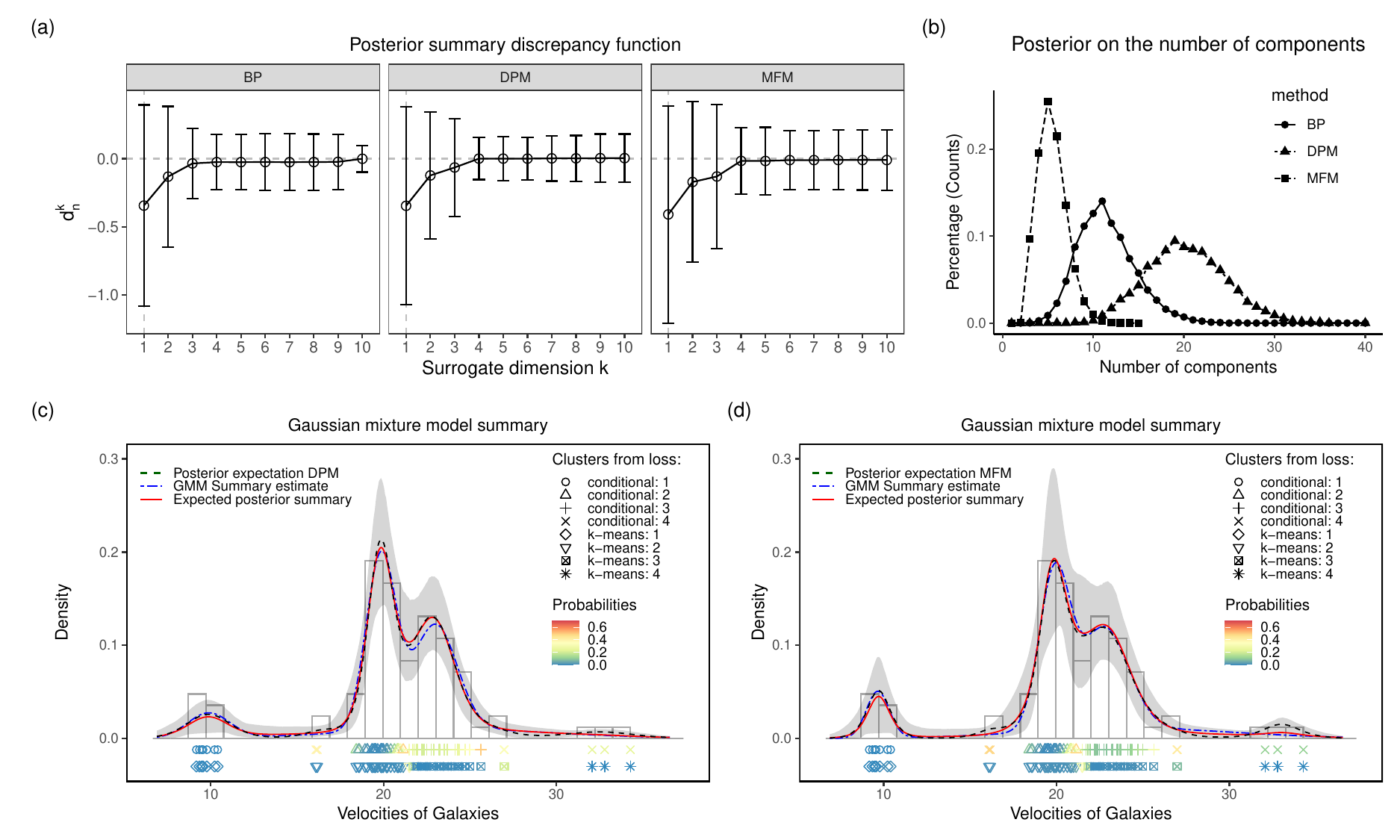}
	\caption{Posterior Summaries for the simulated data: Plot (a) displays the discrepancy functions, $d_{n}^k$, under the respective priors: BP, DPM, and MFM. In panels (c) and (d), the black dashed line shows the expected posterior density under the DPM and MFM models, respectively. The blue line represents the Gaussian summary estimate, $\hat{g}_{\boldsymbol{\gamma}}^{K^*}$, with $K^* = 4$ components. The gray ribbons denote the 95\% credible interval of the posterior summary, while the red line indicates its posterior mean for the five-component Gaussian mixture surrogate. Plot (b) displays the posterior distribution of the number of components under the BP, DPM, and MFM priors.}\label{galaxy_all}
\end{figure*}

We evaluated the data  using the same models as in the previous examples, BP, DPM, and MFM. The posteriors were generated under the same specifications described in Section \ref{simdata}. From the posterior predictive distribution $\tilde{f}$, we drew a sample of size $\tilde{N} = 2000$, for each model. We set $K_{\text{max}} = 10$ and obtained the sequence of density summary estimates using finite mixture models with Gaussian kernels, as described in Algorithm \ref{algo_estimate}. Figure \ref{galaxy_all} (a), interpreted through the heuristic defined in Section \ref{section23}, shows that a Gaussian summary bewteen three and four components, provides a reasonable approximation for the predictive posterior under the BP prior, while four components are needed for the DPM and MFM priors. In these cases, the average discrepancy $\bar{d}_{n}^k$ is close to zero, confirming the adequacy of the approximation.

Given the choices of $K^*$ for the different models, we projected their posteriors via Algorithm \ref{algo2} and obtained the corresponding lower-dimensional approximations. In addition, we applied the conditional probability and $k$-means loss functions, defined in Sections \ref{section51} and \ref{section42}, respectively, to determine cluster allocations and quantify uncertainty.

Figure \ref{galaxy_all} presents the results of our method under the different prior choices. In panels (c) and (d), the galaxy data is shown with the posterior average density estimate under the DPM and MFM priors (black dashed line). The gray ribbons represent the 95\% credible interval of the posterior summary, while the red line indicates its posterior summary mean. In both cases, the expected density lies within the bounds of the summary. Panels (c) and (d) also display the cluster allocations based on conditional probability (top row), as defined in \eqref{clust_alloc}, and $k$-means loss (bottom row), as defined in \eqref{kmeans_aloc}, together with the associated uncertainty measures given in \eqref{Uncert_cond} and \eqref{Uncert_kmeans}, respectively.

As described in Section \ref{section1}, Figures \ref{BP_galaxy_all}(a) and (b) display the galaxy data with the expected posterior density estimate under the BP prior (black dashed line). The gray ribbons represent the 95\% credible interval of the posterior summary, while the red line denotes its posterior mean, $\bar{g}^{K^*}_{\boldsymbol{\gamma}}$. As in the previous cases, the posterior mean density lies within the bounds of the summary in both instances. The figures also include cluster allocation summaries, both estimate and uncertainty allocation, based on conditional probability and $k$-means loss, as defined in Sections \ref{section51} and \ref{section42}, respectively.

In plot (b), Figure \ref{galaxy_all}, we observe the posterior distribution of the number of groups under the BP, DPM, and MFM models.  Additional univariate applications of our method are given in Supplementary Material \ref{appendixB}.

\subsection{Multivariate Models}\label{Mult_models}

In this subsection, we applied our method to models fitted to multivariate data. Our goal is to obtain lower-dimensional density and cluster allocation summaries with uncertainty quantification. The models used in the numerical experiments include the DPM and MFM, introduced in Section \ref{unisim}, and sparse finite mixture models, \cite[SFM;][]{malsiner2016model}.

The SFM is an over-parameterized finite mixture model with a sparsity-inducing prior on the kernel means, leading to optimal choice in the estimation of the number of components and thus enabling model-based clustering. Additional information on the methods, the hyperparameters, and the posterior conditions is given in Supplementary Material \ref{secA1}.

\subsubsection{Simulated Data}\label{sim_mult}

Following the simulation scenario of \cite{miller2018mixture}, we generated $N = 1000$ observations from a three-component bivariate Gaussian mixture, as $\boldsymbol{y}_i\sim \sum_{q= 1}^3 \omega_qN_2(\boldsymbol{y}_i|\boldsymbol{\mu}_q, \boldsymbol{\Sigma}_{q})$ with means $\boldsymbol{\mu}_1 = (4, 4)^T$, $\boldsymbol{\mu}_2 = (7, 4)^T$,  $\boldsymbol{\mu}_3 = (6, 2)^T$, weights $\omega = (0.45, 0.3, 0.25)$, and  with covariance matrices
\begin{equation*}
	\boldsymbol{\Sigma}_1 = \begin{pmatrix} 1 & 0 \\ 0 & 1 \end{pmatrix},
	\boldsymbol{\Sigma}_2 = R \begin{pmatrix} 2.5 & 0 \\ 0 & 0.2 \end{pmatrix} R^T, \text{with } R = \begin{pmatrix} \cos \rho & -\sin \rho \\ \sin \rho & \cos \rho \end{pmatrix},
	\boldsymbol{\Sigma}_3 = \begin{pmatrix} 3 & 0 \\ 0 & 0.1 \end{pmatrix},
\end{equation*}
where $\rho = \pi / 4$.

We initiated our method by obtaining the posterior distributions from the DPM, MFM, and SFM models, from which we generated $\tilde{N} = 2000$ samples from the posterior predictive distribution $\tilde{f}$. We obtained density and cluster summaries using finite mixture model summaries with Gaussian kernels, following the procedures described in Section \ref{section3} and Algorithm \ref{algo_estimate}.

\begin{figure*}[t!]
	\centering
	\includegraphics[width=1\textwidth]{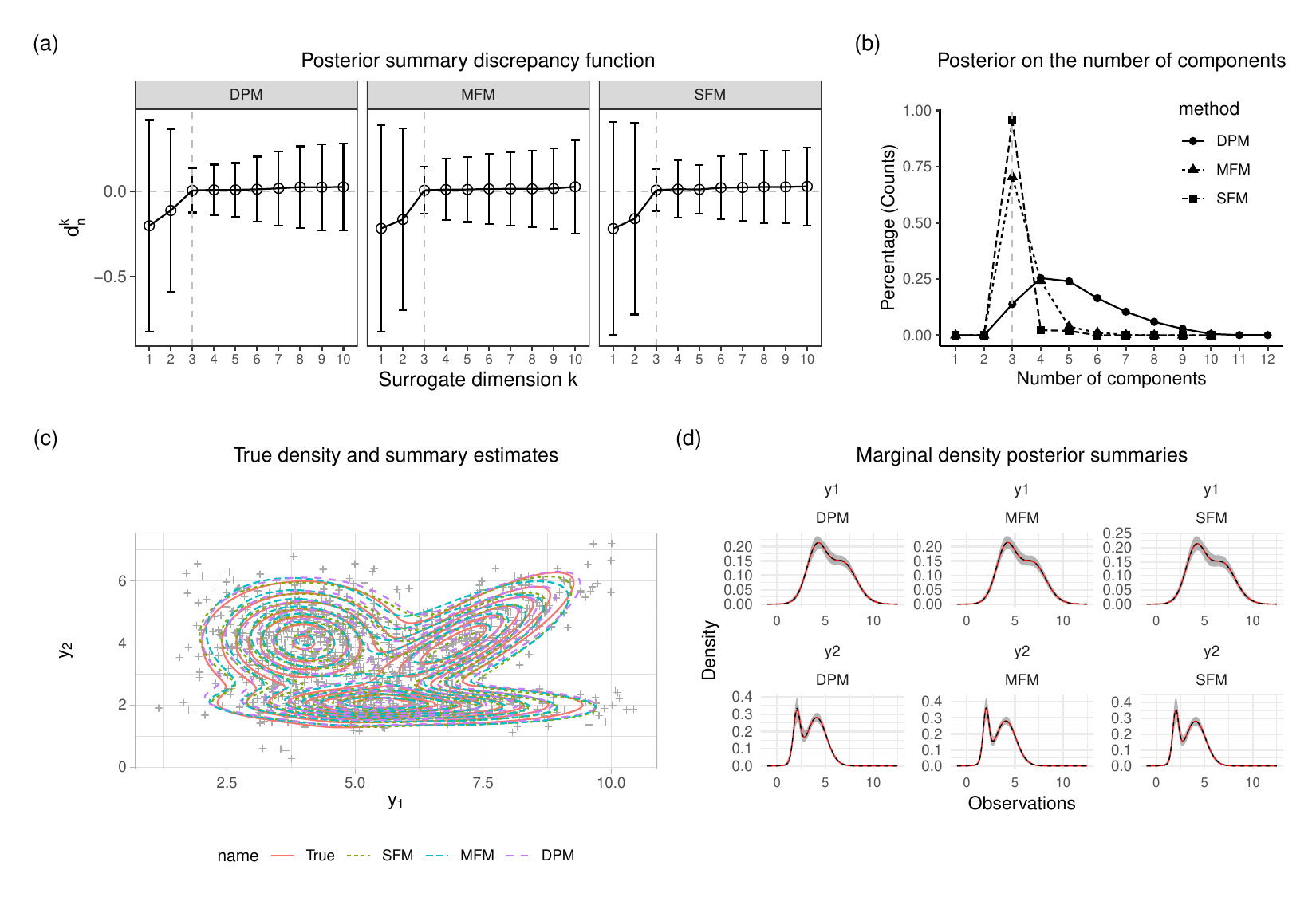}
	\caption{Posterior Summaries for the simulated data: Figure (a) illustrates the discrepancy function, $d_{n}^k$, under different models, DPM, MFM, and SFM, compared with the true number of groups in the data (gray dashed-line). Figure (c) compares the true density with the optimal density summaries estimates, $\hat{g}_{\boldsymbol{\gamma}}^{K^*}$, with $K^* = 3$, under the different models. Figure (d) displays the posterior expected density of the different models (black dashed-line) where the greay ribbon is the 95\% credible interval generated by the posterior summary. The red line indicates the expected posterior summary density, $\bar{g}_{\boldsymbol{\gamma}}^{K^*}$. Figure (b) depicts the posterior distribution of the number of components under the different priors, with the true number of groups (gray dashed line) for comparison.}\label{multi_all}
\end{figure*}

From the discrepancy function in Figure \ref{multi_all}, plot (a), we have that a three-component Gaussian summary, $k = 3$, closely approximates the predictive posterior of the DPM, MFM, and SFM models, as indicated by the average discrepancy function, $\bar{d}_{n}^k$, being near zero. Variability $\text{sd}(d_{n}^k)$ increases noticeably for summaries with fewer, $k < 3$, or more, $k > 3$, components, suggesting that $k = 3$ provides the best overall approximation. Figure \ref{multi_all}, plot (b), displays the posterior distribution of the number of components across the three models to the true number of groups. We then applied Algorithm \ref{algo2} and obtained the posterior summaries for $K^* = 3$.

Figure \ref{multi_all} (c) compares the true density with the summary estimates obtained from a Gaussian mixture model summary $\hat{g}_{\boldsymbol{\gamma}}^{K^*}$, with $K^* = 3$ components, under the different models. Figure \ref{multi_all} (d) displays the 95\% credible interval (gray ribbon) of the posterior summary under the different priors. The red line is the expected value of the posterior summary $\bar{g}_{\boldsymbol{\gamma}}^{K^*}$, compared to the true density (black line). 

\begin{figure}[t!]
	\centering
	\includegraphics[width=1\linewidth]{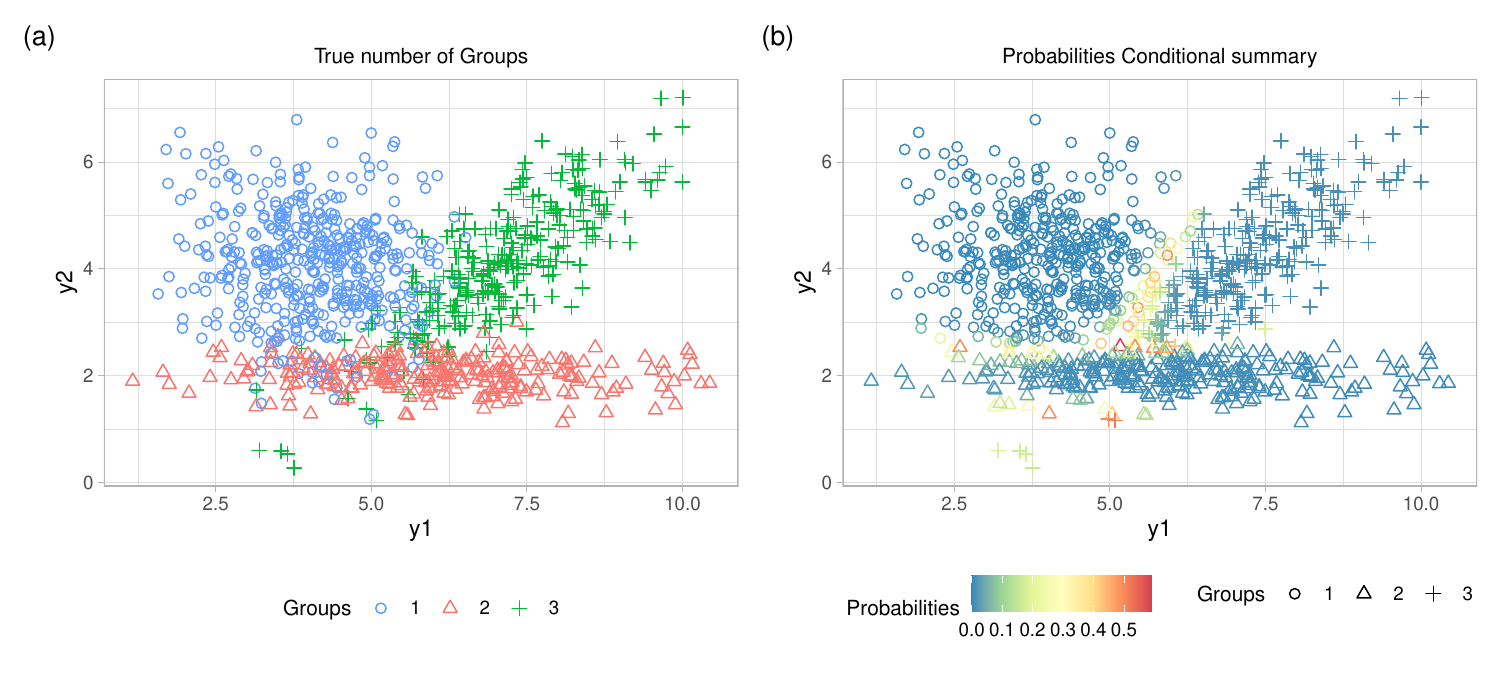}
	\caption{Figure (a) shows the true cluster allocation for the simulated dataset, while Figure (b) presents the summary cluster allocations defined in \eqref{clust_alloc} with $K^* = 3$ groups and uncertainty quantification, as described in \eqref{Uncert_cond}, using the conditional probability allocation under the DPM model.}\label{cluster_mult1}
\end{figure}

Figure \ref{cluster_mult1} (b) displays the cluster allocations and uncertainty estimates obtained using the conditional probability function, defined in Section \ref{section51}, under the DPM model. The number of the summary data clusters was set to $K^*=3$ where the cluster summary estimates were obtained using \eqref{clust_alloc}, with allocation uncertainty quantified by \eqref{Uncert_cond}. Figure \ref{cluster_mult1} (a) displays the true cluster allocation.

We evaluated clustering performance for the different models relative to their summaries cluster estimates using the adjusted Rand index (ARI), which measures the agreement between true and estimated cluster memberships. Basically, an ARI near zero corresponds to two independent partitions, while an ARI near one indicates perfect agreement. We also report the classification error (err) for each model, where values close to zero reflect good classification performance.

\begin{table}[h]
	\caption{Original and summary clustering performance of the DPM, MFM, and SFM models, evaluated using the Adjusted Rand Index (ARI) and classification error (err). Particular emphasis is placed on the cluster summary estimate with conditional probability allocation \eqref{clust_alloc}. We compared the results to those obtained from the \texttt{mclust} package. We highlighted the model with the best performance.}\label{tab:clust_results}
	\begin{tabular*}{\textwidth}{@{\extracolsep\fill}lccccccc}
		\toprule%
		& \multicolumn{3}{@{}c@{}}{cluster summary estimate} & \multicolumn{4}{@{}c@{}}{} \\\cmidrule{2-4}%
		Method & DPM & MFM & SFM& DPM & MFM& SFM & \texttt{mclust} \\
		\midrule
		ARI &  \bf{0.779}      & 0.758 & 0.763  & 0.499     & 0.676    & 0.751 & 0.760\\
		err  & \bf{0.079}      & 0.087     &0.085  & 0.242   & 0.120   & 0.089  &  0.086 \\
		\botrule
	\end{tabular*}
\end{table}

Table \ref{tab:clust_results} displays the ARI index under the different models, with summary cluster estimates, generated under our method, compared to cluster allocation generated by the original methods, and under \texttt{mclust}. We highlighted the model with the best performance.

\subsubsection{Thyroid Data}\label{thyroid}

The thyroid dataset is a widely used benchmark for analyzing multivariate normal mixtures \citep{scrucca2023model}. It consists of five laboratory test variables and a categorical variable indicating the diagnostic outcome for a total of 215 patients. The diagnostic outcome contains three possible scenarios that are the clusters of interest. 

We evaluated the data using the same models, DPM, MFM, and SFM, and hyperparameter specifications as in Section \ref{sim_mult}. We generated a sample of size $\tilde{N}= 2000$ from the posterior predictive distribution $\tilde{f}$ for all models. We constructed summary estimates using finite mixture models summaries with Gaussian kernels, as described in Section \ref{num_summary}, in all cases. The number of components in the finite mixture model summary were selected based on the discrepancy function, $d_{n}^k$, and the heuristic described in Section \ref{section23}. 

\begin{figure}[t!]
	\centering
	\includegraphics[width=1\linewidth]{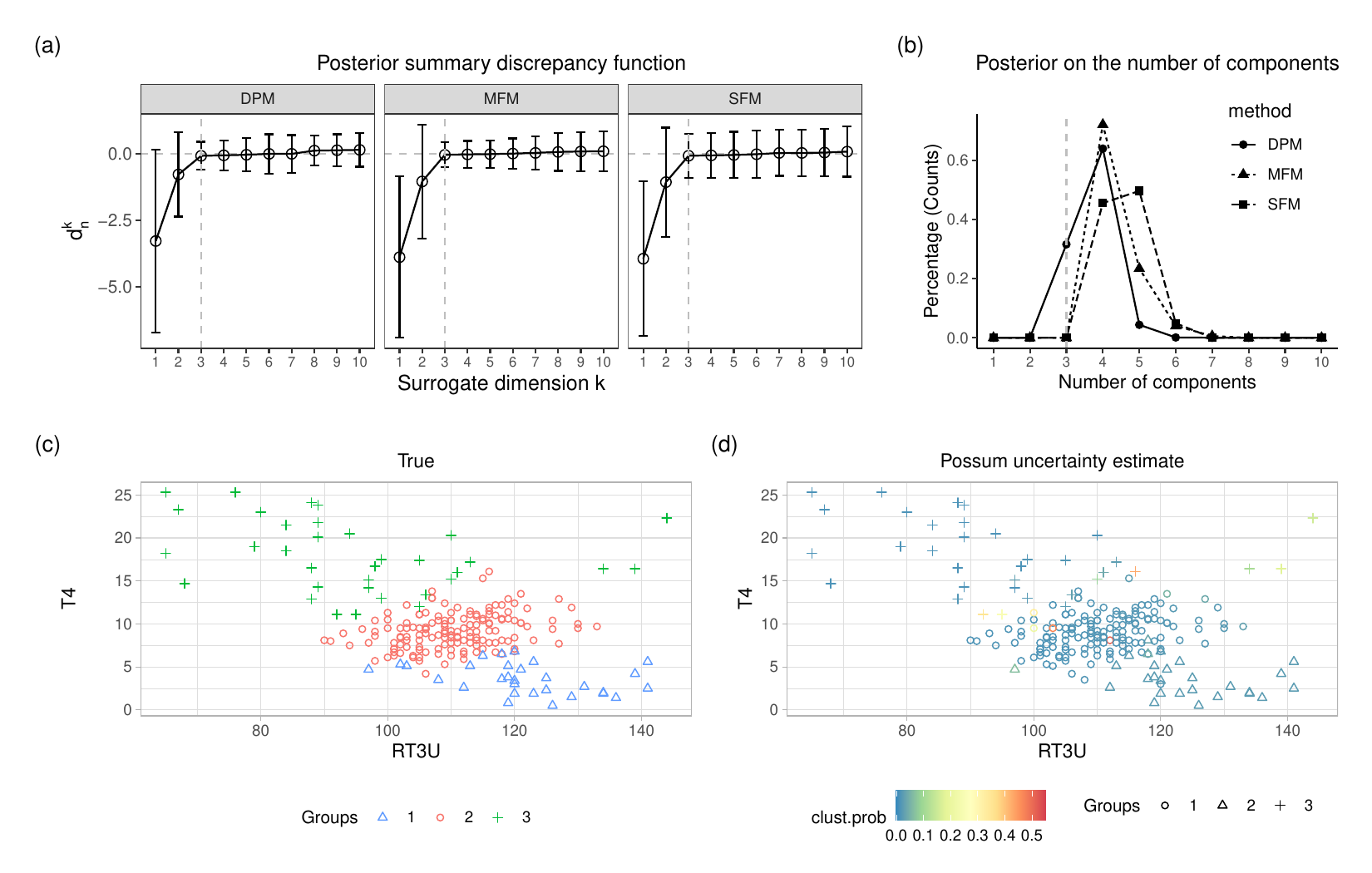}
	\caption{Posterior Summaries Thyroid data: Figure (a) illustrates the discrepancy function, $d_{n}^k$, under different models, DPM, MFM, and SFM, compared with the true number of groups in the data (gray dashed-line). Figure (d) illustrates the optimal cluster allocation summary, with $K^* = 3$ groups, with the uncertainty quantification under the conditional probability loss, for the SFM model. Figure (c) displays the true cluster allocation. Figure (b) displays the original  posterior distribution of the number of components under the different models.}\label{thyroid_plots}
\end{figure}

In Figure \ref{thyroid_plots}, plot (a), the discrepancy plots reveal that a Gaussian summary with three, $k = 3$, provides a reasonable approximation for the predictive posterior of the DPM, MFM, and SFM models. This conclusion is supported by the average value of the discrepancy function, $\bar{d}_{n}^k$, defined in \eqref{approx}, being near zero in all cases. Additionally, the variability associated with the discrepancy function, $\text{sd}( d_{n}^k)$, increases substantially for summaries with $k< 3$ components. 

We generated cluster allocations and uncertainty estimates for the SFM model with $K^* = 3$, using conditional probability allocation summary, defined in Section \ref{section51}. Figure \ref{thyroid_plots} (c) presents the true cluster allocation with respect to the variables T4 and RT3U. For more information on the interpretation of these variables, we refer to \cite{scrucca2023model}. Plot (d) shows the optimal cluster allocation summary with $K^* = 3$ components, generated under the SFM posterior predictive distribution, using the conditional probability allocation loss defined in Section \ref{section51}, while Figure \ref{thyroid_plots} (b) displays the posterior distribution of the number of components from the original posteriors.

We performed an additional evaluation of the cluster summary estimates generated under the SFM model. The ARI for the cluster allocation under the original SFM posterior is 0.88, with a classification error of 0.06, compared to 0.91 and 0.028 for the optimal clustering summary from the SFM posterior. Similarly, the \texttt{mclust} method yields an ARI of 0.877 and a classification error of 0.037.

\section{Concluding Remarks}\label{section7}

In this paper, we proposed a novel posterior summarization method designed to approximate overparametrized and nonparametric Bayesian models. The primary objective of our approach is to preserve interpretability while maintaining model accuracy. The procedure involves three main steps. First, the data is fitted using overparameterized or nonparametric models. In the second step, a decision-theoretic framework is employed to project the posterior distribution to a finite mixture model summary estimate. We introduced a discrepancy function and plot to quantify the trade-off between accuracy and parsimony, guiding the selection of the summary that best approximates the original model. Finally, in the third step, the original posterior distribution is projected onto the summary estimate, yielding a lower-dimensional posterior summary equipped with uncertainty quantification. The effectiveness of the proposed approach in generating interpretable density and cluster summaries from various Bayesian overparameterized and nonparametric models was demonstrated through a series of numerical simulations and applications.

Despite the promising results, several challenges remain. In our experiments, we observed that the effectiveness of our method can be compromised by unsual observations in the posterior predictive samples. To address this issue, robust Bayesian sampling techniques may be employed to mitigate the influence of such samples. Another bottleneck of our method is label switching from the posterior cluster summarie. Different methods have been proposed to approach this issue and can be adopted for our method \citep{fruhwirth2011dealing,malsiner2016model}.  

While the KL divergence property, highlighted in \ref{loss}, has appealing theoretical properties and is widely used for comparing probability distributions, it is also known to be sensitive to discrepancies in the tails, which can pose challenges for estimation and numerical stability. In this regard, exploring other loss functions, such as the Expected Misclassification Rate (EMR) loss, as in applied by \cite{adrian2025scenario}, could lead to more robust or computationally efficient procedures. The EMR loss, for instance, may offer advantages in terms of approximation via Monte Carlo methods. Integrating such alternatives into our framework could provide further insights and improved performance in practical applications.

Future research directions include incorporating a penalty function in the summary loss in \eqref{eq:min_lamb} to induce sparsity in high-dimensional settings \citep{pan2007penalized}. Exploring alternative kernels to address potential model misspecification and theoretically analyzing the method's convergence and robustness under different prior specifications are also potential directions. Additionally, further research could explore variational Bayes methods \citep{murphy2012machine} to improve computational efficiency and extend our method to posterior summaries based on predictive distributions, including Martingale Posteriors \citep{fong2023martingale, rodriguez2025martingale}.

\bibliography{sn-bibliography}

\appendix
\begin{appendices}

\section{Model specification}\label{secA1}

\subsubsection{Random Bernstein polynomial  model}

The random Bernstein polynomial \cite[BP;][]{petrone1999random}, model is widely applied in density estimation for data with support on the unit interval. Particularly, is useful when other methods that are susceptible to boundary effects fail. With a rather simple variable transformation, this model can be extended to unbounded data. In a nutshell, the BP prior is defined by the Bernstein polynomial, where the number of terms is random and the data partition, on which the polynomial is fit, is defined by a Dirichlet process \citep{ferguson1973bayesian,ferguson1983bayesian}. The Bernstein polynomial, associated with a cumulative distribution function \(G\) on \([0,1]\), is defined as $B(y|q,G) = \sum_{j  = 0}^q G(\frac{j}{q})\binom{q}{j}y^j(1-y)^{q-j}$, with \(q\in \mathbb{Z}_{+}\), with corresponding density $b(y|q,G) = \sum_{j = 1}^qG\left(\frac{j-1}{q},\frac{j}{q}\right]\text{Beta}(y|j,q-j+1)$.
If \(G\) has a continuous density \(f\), then \(b(y| q, G)\) converges uniformly to \(f\) as \(q\) grows to infinity. By letting $G \sim\mathrm{DP}(\alpha, G_{0})$, a Dirichlet process, and by letting a prior on \(q\) on the naturals, results in a random Bernstein Dirichlet process, which can be viewed as a smoothed Dirichlet process \citep{petrone1999random}. Key properties such as estimation, consistency, and convergence rates of the BP model have been explored by \cite{petrone2002consistency}. Throughout the numerical applications of this paper, we set the base measure $G_0$ as a Beta distribution, $\text{Beta}(1,1)$, we let a $\text{Gamma}(2.01,0.1)$ distribution prior on $\alpha$, and $q$ to have a discrete uniform distribution prior $\text{Uniform}\{1,\dots,1000\}$. For the simulations and applications in Section \ref{unisim} we used the \texttt{Dppackage} library from \citep{JSSv040i05}, where we generated a posterior of size 6,000, with a burn-in of 1,000, in which we applied a thinning of size 10.

\subsubsection{Dirichlet process mixture model}

The Dirichlet process mixture model (DPM) provides a nonparametric framework for clustering and density estimation  \citep{ferguson1983bayesian, escobar1995bayesian}. As seen in Section \ref{section1}, the flexibility of the DPM stems from the model's ability to adapt to an unknown number of clusters in the data and to its computational efficiency, as highlighted by \cite{Maceachern01061998}. In this paper, the mixture model is taken to be Gaussian, $f(\cdot | \boldsymbol{\theta})  = \frac{1}{\sqrt{2\pi\sigma^2}}\exp(-(\cdot-\mu)^2/2\sigma^2)$ where  $\boldsymbol{\theta} = (\mu, \sigma^2)$, with a conjugate Normal-Gamma prior for the base measure $G_{0}\!\left(\boldsymbol{\theta} | \kappa\right) = N(\mu| \mu_{0},\frac{\sigma^{2}}{k_{0}}) \text{Gamma}\left(\sigma^{-2} |\, \alpha_{0},\beta_{0}\right)$, parameterized by 
$\kappa=(\mu_{0},k_{0},\alpha_{0},\beta_{0})$. We adopted the hyperparameter configuration
$\mu_{0}=\texttt{median}(\boldsymbol{y})$, $k_0 = 1/5$, $\alpha_{0}=2$, $\beta_{0}=1$, and a $\text{Gamma}(2, 4)$ distribution prior for the concentration parameter $\alpha$. The posterior is generated using the R package \texttt{dirichletprocess} \citep{Ross2023dirichletprocess}.
For the simulations and applications in Section \ref{unisim}, we generated a posterior sample of size 6,000 with a burn-in of 1,000.

For the multivariate DPM, we have the model $f(\cdot|\boldsymbol{\theta})$,  with $\boldsymbol{\theta} = (\boldsymbol{\mu},\boldsymbol{\Sigma})$, where $\boldsymbol{\mu}\in \mathbb{R}^{d\times 1}$, and $\boldsymbol{\Sigma}\in \mathbb{R}^{d\times d}$, we have the base measure $G_{0}(\boldsymbol{\theta}|\mu_1,\kappa_1,\nu_0,T_0) = N(\boldsymbol{\mu}|\mu_1,\kappa_{1}^{-1}\boldsymbol{\Sigma})W_{d}(\boldsymbol{\Sigma} | \nu_0,T_0)$, where $W_{d}(\boldsymbol{\Sigma} | \nu_0,T_0)$ is the Wishart distribution with $\nu_0$ degrees of freedom, and where $T_0$ is a symetric positive definite matrix. In the simulations in Section \ref{sim_mult}, we have $\kappa_1 = 1$, $\nu_0 = 2$, and $T_{0} = \text{Cov}(\boldsymbol{y})$, the sample covariance matrix, and a $\text{Gamma}(2, 4)$ distribution prior for the concentration parameter $\alpha$. For the application in Section \ref{thyroid}, we let $\nu_0 = 5$. In both cases, for the simulations and applications, we generate 2,000 posterior samples, with a burn-in of 1,000. 

\subsubsection{Mixtues of Finite Mixtures}

As presented in Section \ref{section1}, finite mixture models assume that the data are generated from a mixture of component distributions belonging to a well-defined parametric family. One of the issues of finite mixture models is that the number of components has to be fixed in advance, before inference is made. A natural extension of finite mixture models is to treat the number of components as an unknown parameter in which we can associate a distribution, and as consequently a prior \citep{richardson1997bayesian,miller2018mixture}. As a result we have the mixture of finite mixtures model (MFM). In this paper, the kernels are Gaussian distributions, $f(\cdot | \boldsymbol{\theta}_q) = \frac{1}{\sqrt{2\pi\sigma^2}}\exp(-(\cdot-\mu_q)^2/2\sigma_{q}^2)$, where $\boldsymbol{\theta}_q = (\mu_q, \sigma_{q}^2)$,  independent priors defined as $\mu_q \sim N\bigl(\mu|\mu_0,\sigma_0^2\bigr),  \sigma^{-2} \sim \mathrm{Gamma}(\sigma^{-2}|a,b)$.
A hyperprior is placed on \(b\) via $b \sim \mathrm{Gamma}(b|a_0,b_0)$. For the weights, we have a Dirichlet distribution $(\omega_1,\dots, \omega_q)\sim\text{Dir}(\omega_1,\dots, \omega_q|1)$.  

We used the out-of-the-box hyperparameters configuration, from the Julia implementation\footnote{\url{https://github.com/jwmi/BayesianMixtures.jl}} by  \cite{miller2018mixture}. For additional details on the choice of hyperparameters and the sampling procedure we refer to \cite{miller2018mixture}. For MFM in Section \ref{unisim}, we ran a total of 100,000 iterations per dataset; we discarded the first 10,000 iterations as burn-in and kept 5,000 by thinning.

For the multivariate MFM, in Section \ref{Mult_models}, we again used a Gaussian kernels, and independent priors for the location and the covariance matrix. We take $\boldsymbol{\mu} \sim N(\boldsymbol{\mu} |\hat{\mu},\hat{C})$, $\boldsymbol{\Sigma}^{-1}\sim W_{d}(\boldsymbol{\Sigma}|\nu,V)$, where $\hat{\mu}$ is the sample mean, and $\hat{C}= \text{Cov}(\boldsymbol{y})$ is the sample covariance matrix, $\nu = 2$, and $V = \hat{C}/\nu$. We use $K \sim \text{Geometric}(0.1)$, and a Dirichlet distribution, $(\omega_1,\dots,\omega_q) \sim \text{Dir}(\omega_1,\dots, \omega_q|1)$, for the weights. We generate a posterior sample of size 60,000, with 5,000 as burn in, where 1,000 are kept. 

\subsection{Sparse Finite Mixture Models}

For the Sparse Finite Mixture Model \cite[SFM;][]{malsiner2016model}, we have a Gaussian mixtute model, where we fix the number of components before hand as $K\in \mathbb{Z}_{+}$, $\boldsymbol{\mu}_k\sim N(\boldsymbol{\mu}_k|b_0,B_0)$, $b_0 = \texttt{median}(\boldsymbol{y})$, $B_0 = R_0$, $R_0 = \text{diag}(R_{1}^2, \dots, R_{d}^2)$, where $R_j$, for  $j = 1,\dots,d$, are vectors containing the range of each column of the data $\boldsymbol{y}$. We let  $(\omega_1,\dots, \omega_K)\sim\text{Dir}(\omega_1,\dots, \omega_K|0.01)$, for the weights. For the SFM in Section \ref{Mult_models}, we generate a posterior sample of size 2,000, with 200 as burn-in, where 1,000 are kept. 

\section{Numerical applications}\label{appendixB}

In this supplementary material, we present the results from our method applied to the remaining benchmark datasets, acidity and enzyme, from \cite{richardson1997bayesian}. 

The acidity data concerns the acidity index measured in a sample of 155 lakes in north-central Wisconsin. This index describes the capacity of a lake to absorb acid. Low values in the data may indicate a loss of biological resources. This dataset was initially analyzed by \cite{crawford1992modeling} and later by  \cite{richardson1997bayesian}, who considered three to five components relevant. The enzyme data concern the distribution of a specific blood enzyme involved in the metabolism of carcinogenic substances, measured in 245 individuals. The analysis focuses on identifying two subgroups of slow and fast metabolizers, which serve as markers in the general population. In \cite{richardson1997bayesian}, three to five components were also identified as relevant.

We use the same Bayesian models as those employed in the univariate numerical applications in Section \ref{section6}. These models include the Bernstein polynomial prior (BP), the Dirichlet process mixture (DPM), and mixtures of finite mixtures (MFM). The posteriors under these choices were generated under the same specifications as defined in Supplementary Material \ref{secA1}. We generated summaries with uncertainty quantification using finite mixture models, as described in Section \ref{section3}. From the posterior predictive distribution $\tilde{f}$, we generated a sample of size $\tilde{N}= 2000$. We generated the density summary estimates with uncertainty quantification using finite mixture models, as described in Algorithms \ref{algo_estimate} and \ref{algo2}, respectively. We implemented the conditional probability allocation and k-means loss, defined in Sections \ref{section51} and \ref{section42}, respectively, to determine cluster allocations and uncertainty estimates. 



\subsubsection{Acidity Data}

\begin{figure}[t!]
	\centering
	\includegraphics[width=1\linewidth]{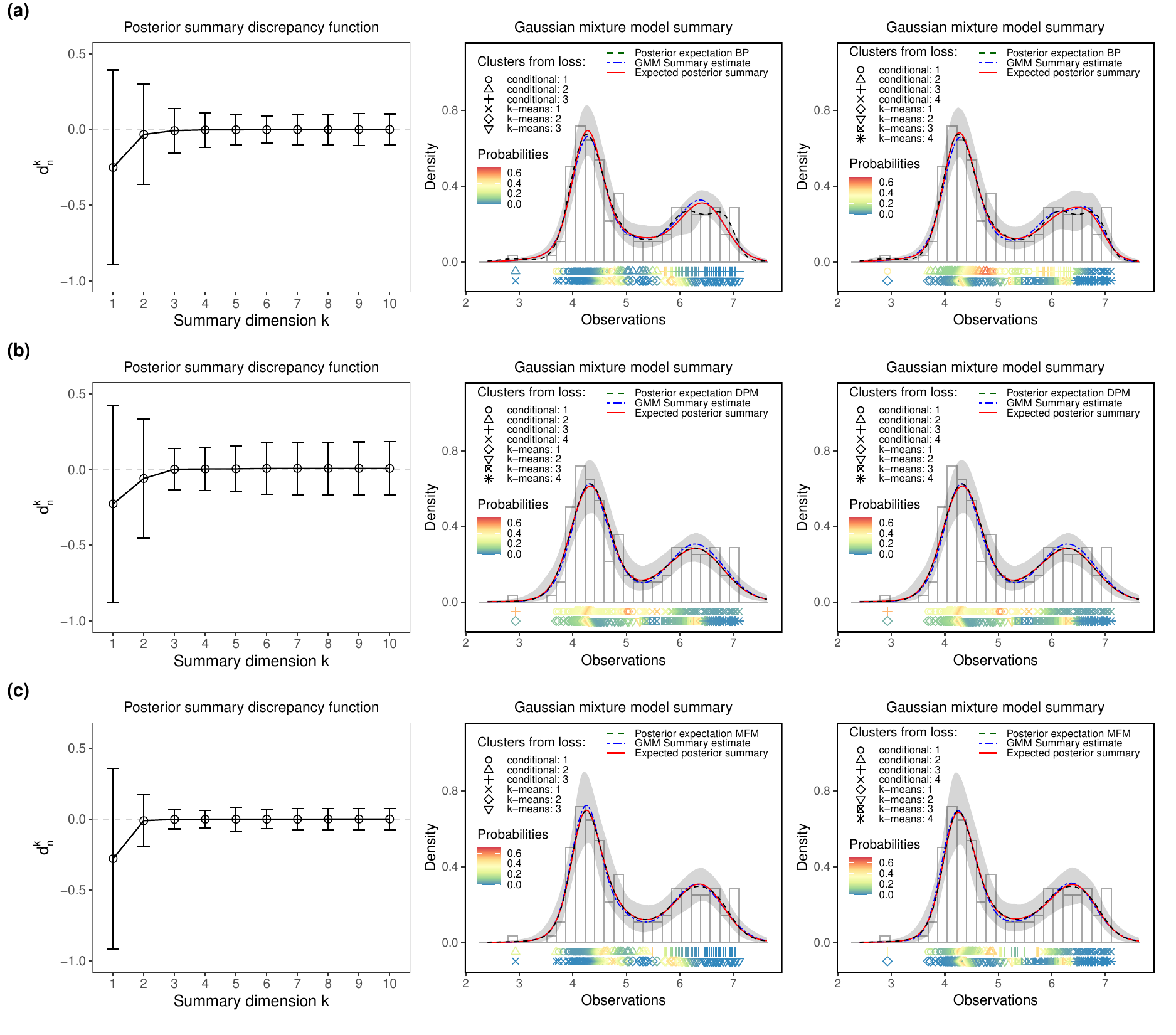}
	\caption{Posterior Summaries for the acidity data: In Figures (a), (b), and (c), the left plots display the discrepancy function, $d_{n}^k$, for the BP, DPM, and MFM models, respectively. The center and right plots display the acidity data, with the posterior expectation under different models. The plots also present summary estimates, with $K^* = 3$, and $K^* = 4$ components (blue dot-dashed line). Additionally, it displays cluster allocations based on conditional probability (row 1) and k-means loss (row 2), along with the associated uncertainty and quantification for the allocations.}\label{acidity1}
\end{figure}

Figure \ref{acidity1} displays the results of the proposed method applied to the acidity data under BP, DPM, and MFM. In Figure \ref{acidity1}, plot (a), (b), and (c), the left plot displays the discrepancy plots $d_{n}^k$, introduced in Section \ref{section23}, and defined in \eqref{eq:6}. The plots reveal that a GMM summary, with three components, $k = 3$, provides a reasonable approximation for the predictive posterior distribution $\tilde{f}$, defined in \ref{section21}, for all evaluated models, since the average value of the discrepancy function $\bar{d}_{n}^k$, in \eqref{approx}, is close to zero. However, for the BP model, there's an increase in uncertainty for this approximation. The central and right plots of Figures \ref{acidity1}, (a), (b), and (c) display the acidity data with the expected posterior density for the different models (black-dashed line). The plots also show summary estimates with $K^* = 3$ and $K^* = 4$, (blue dot-dashed line), on the center and right plots, respectively. The gray ribbon is the 95\% credible interval of the posterior summary, where the red line indicates the posterior predictive average. Additionally, we display cluster allocation summaries based on the conditional (row 1), defined in Section \ref{section51}, and k-means (row 2), defined in Section \ref{section42}. We also display the uncertainty on the cluster allocation for the conditional and k-means loss defined in \eqref{Uncert_cond} and \eqref{Uncert_kmeans}, respectively. 

In Figure \ref{post_acidity_enzyme} (a), the posterior distributions of the number of clusters under the BP, DPM, and MFM models are shown. The posterior distribution under the BP model concentrates around 13 components, around five components under the DPM model, and around three components under the MFM model.

\subsubsection{Enzyme Data}

\begin{figure}[t!]
	\centering
	\includegraphics[width=1\linewidth]{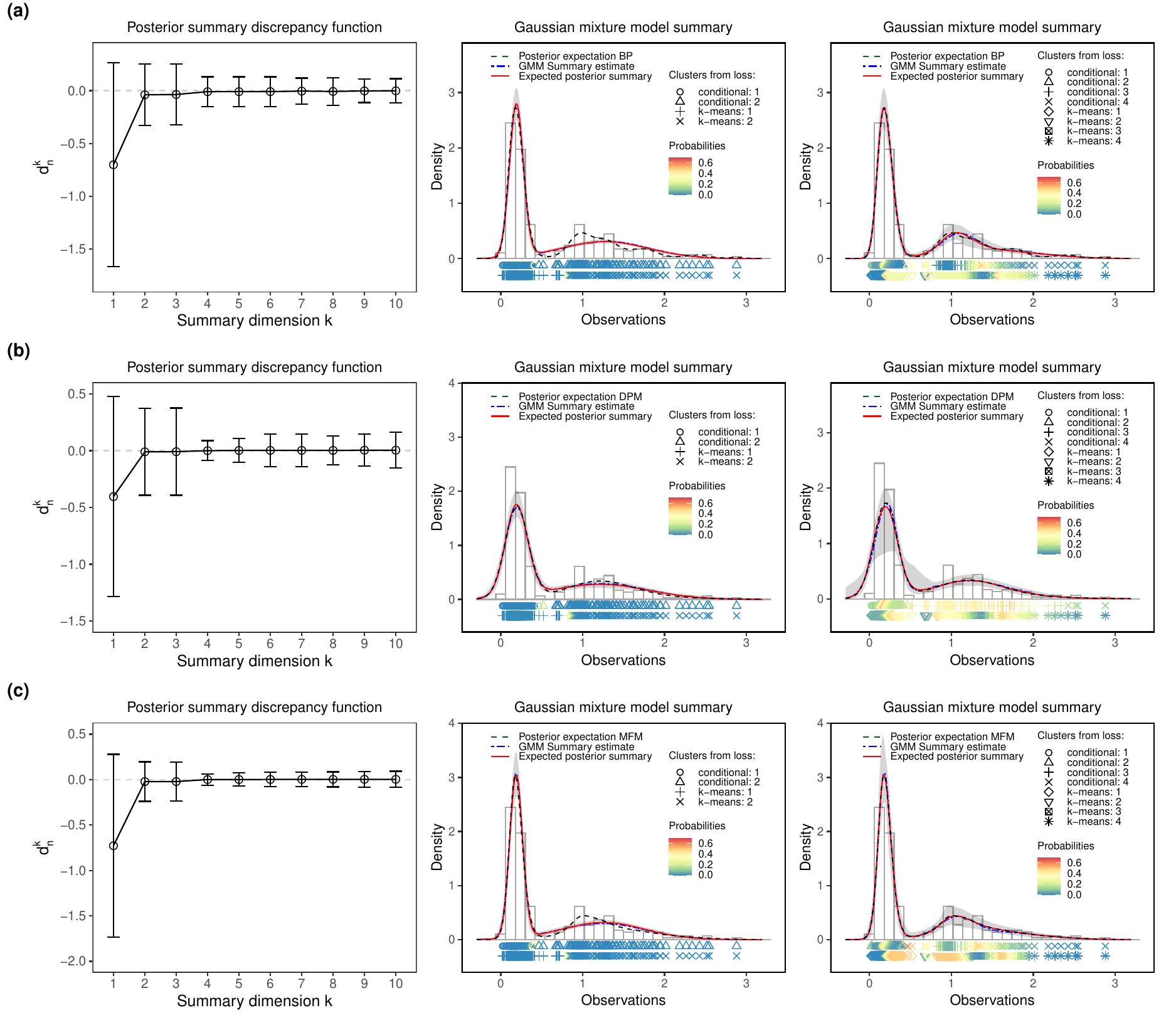}
	\caption{Posterior summaries for the enzyme data: In Figures (a), (b), and (c), the left plots display the discrepancy function, $d_{n}^k$, for the BP, DPM, and MFM models, respectively. The center and right plots display the enzyme data, with the posterior expectation under different models. The plots also present summary estimates, with $K^ = 2$ and $K^* = 4$ components (blue dot-dashed line). Additionally, they display cluster allocations based on conditional probability (line 1) and k-means loss (line 2), along with the associated uncertainty for the allocations.}\label{enzyme1}
\end{figure}

Figure \ref{enzyme1} shows the results of our method on the enzyme data. In plots (a), (b), and (c), the left plots display the discrepancy function $d_{n}^k$, where a summary estimates with $k = 2$ components provides a reasonable approximation to the posterior predictive distribution under the BP, DPM, and MFM models. This is supported by the fact that the average value of the discrepancy function $\bar{d}_{n}^k$ is close to zero for all models. The center plots show the summary estimates with $K^* = 2$ and $K^* = 4$ components (blue dot-dashed line), compared with the expected posterior of each model (black dashed line). The gray ribbons represent the 95\% credible intervals of the posterior summaries, with the red line indicating the posterior average. Cluster summaries for the conditional (row 1) and k-means (row 2) losses are also displayed, with associated uncertainty estimates.

\begin{figure}[t!]
	\centering
	\includegraphics[width=1\linewidth]{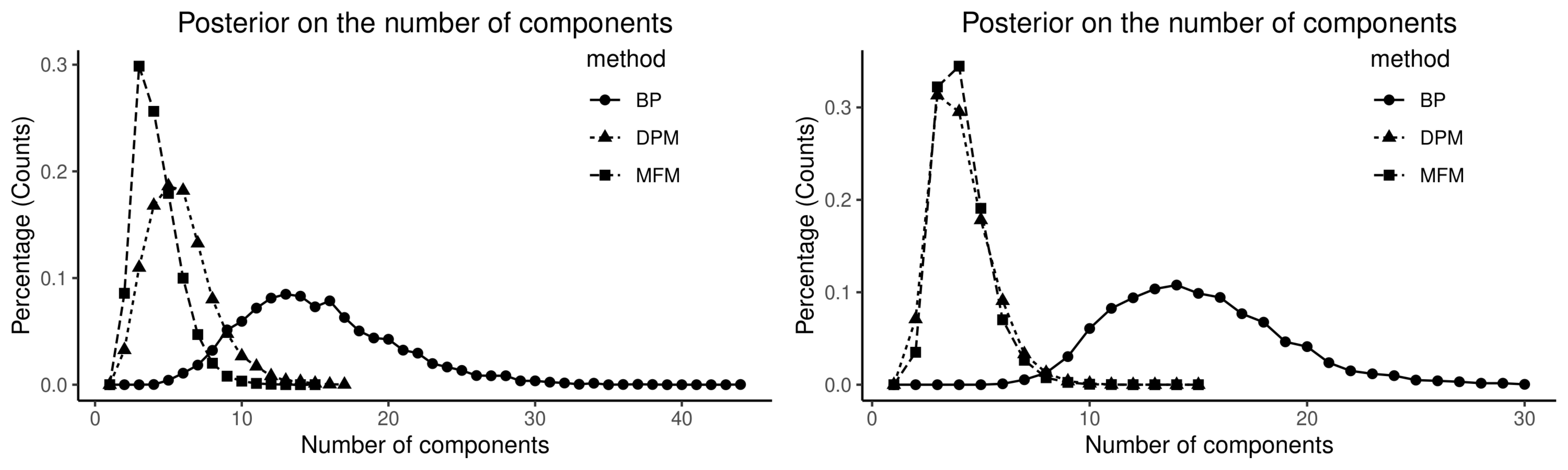}
	\caption{Posterior probabilities for the number of components under the BP, DPM, and MFM priors for the Acidity data (left) and Enzyme data (right), respectively.}\label{post_acidity_enzyme}
\end{figure}

In Figure \ref{post_acidity_enzyme} (b), we observe the posterior distribution of the number of clusters under the BP, DPM, and MFM models. Under the BP model, the posterior concentrates around 14 components, whereas for the DPM and MFM models, they concentrate between three and four components.

\end{appendices}



\end{document}